\documentclass{emulateapj2}

\newcommand{\msun}{M$_{\sun}$}
\newcommand{\mjup}{M$_{Jup}$}
\newcommand{\rjup}{R$_{Jup}$}
\newcommand{\ldl}{$\lambda/{\Delta}{\lambda}$}
\newcommand{\logg}{$\log{g}$}
\newcommand{\teff}{T$_{\rm eff}$}
\newcommand{\lbol}{$\log_{10}{{\rm L}_{bol}/{\rm L}_{\sun}}$}
\newcommand{\lhalbol}{$\log_{10}{{\rm L}_{H\alpha}/{\rm L}_{bol}}$}

\newcommand{\meth}{CH$_4$}
\newcommand{\wat}{H$_2$O}

\newcommand{\kms}{km~s$^{-1}$}
\newcommand{\cmss}{cm~s$^{-2}$}
\newcommand{\vtan}{$V_{\rm tan}$}
\newcommand{\vrad}{$V_{\rm rad}$}

\newcommand{\name}{2MASS~J13153094$-$2649513}
\newcommand{\namesh}{2MASS~J1315$-$2649}

\slugcomment{Accepted for Publication to ApJ }

\shorttitle{A Companion to 2MASS~J1315$-$2649}
\shortauthors{Burgasser et al.}

\begin{document}

\title{The Hyperactive L Dwarf 2MASS~J13153094$-$2649513: Continued Emission and a Brown Dwarf Companion\footnote{Data presented herein were obtained at the W.\ M.\ Keck Observatory, which is operated as a scientific partnership among the California Institute of Technology, the University of California and the National Aeronautics and Space Administration, and made possible by the generous financial support of the W.\ M.\ Keck Foundation; and with the 6.5-m Magellan Telescopes located at Las Campanas Observatory, Chile.}}

\author{
Adam J.\ Burgasser\altaffilmark{a,b,c,d},
Breann N.\ Sitarski\altaffilmark{a,e},
Christopher R.\ Gelino\altaffilmark{f},
Sarah E.\ Logsdon\altaffilmark{a},
and
Marshall D.\ Perrin\altaffilmark{g}
}

\altaffiltext{a}{Center for Astrophysics and Space Science, University of California San Diego, La Jolla, CA 92093, USA; aburgasser@ucsd.edu}
\altaffiltext{b}{Massachusetts Institute of Technology, Kavli Institute for Astrophysics and Space Research, 77 Massachusetts Avenue, Cambridge, MA 02139, USA}
\altaffiltext{c}{Hellman Fellow}
\altaffiltext{d}{Visiting Astronomer at the Infrared Telescope Facility, which is operated by the University of Hawaii under Cooperative Agreement no. NNX-08AE38A with the National Aeronautics and Space Administration, Science Mission Directorate, Planetary Astronomy Program.}
\altaffiltext{e}{Department of Physics and Astronomy, UCLA, Los  Angeles, CA 90095-1562, USA}
\altaffiltext{f}{Infrared Processing and Analysis Center, MC 100-22, California Institute of Technology, Pasadena, CA 91125, USA}
\altaffiltext{g}{Space Telescope Science Institute, 3700 San Martin Dr, Baltimore, MD 21218, USA}

\begin{abstract}
We report new observations of the unusually active, high proper motion L5e dwarf 2MASS J13153094$-$2649513.  
Optical spectroscopy with Magellan/MagE reveals persistent nonthermal emission, with narrow H~I Balmer, Na~I and K~I lines all observed in emission.  
Low-resolution near-infrared spectroscopy with IRTF/SpeX indicates the presence of a low-temperature companion, which is resolved through multi-epoch laser guide star adaptive optics 
imaging at Keck.  The comoving companion is separated by 338$\pm$4~mas, and its relative brightness ($\Delta{K_s}$ = 5.09$\pm$0.10) makes this system the second most 
extreme flux ratio very low-mass binary identified to date.
Resolved near-infrared spectroscopy with Keck/OSIRIS identifies this companion 
as a T7 dwarf.
The absence of Li~I absorption in combined-light optical spectroscopy 
constrains the system age to $\gtrsim$0.8--1.0~Gyr, while the system's kinematics
and unusually low mass ratio (M$_2$/M$_1$ = 0.3--0.6) suggests that it is even older.
A coevality test of the components also indicates an older age, but reveals discrepancies between evolutionary and atmosphere model fits of the secondary which are likely attributable to poor reproduction of its near-infrared spectrum. 
With a projected separation of 6.6$\pm$0.9~AU, the {\namesh} system is 
too widely separated for mass exchange or magnetospheric interactions to be powering its persistent nonthermal emission.  Rather, the emission is probably chromospheric in nature, signaling an inversion in the age-activity relation in which strong magnetic fields are maintained by relatively old and massive ultracool dwarfs. 
\end{abstract}


\section{Introduction}

Nonthermal emission is commonly observed among the lowest-mass stars, traced by optical line (e.g., Ca II, H I sequence),  X-ray, UV and radio emission.  This emission can be both persistent (quiescent) and eruptive, with short-duration flares from M dwarfs occurring at a rate of roughly 3\%  \citep{2010AJ....140.1402H}.  The incidence and strength of quiescent magnetic activity as traced by the H$\alpha$ line reaches $\gtrsim$80\% and {\lhalbol} $\approx$ $-$4, respectively, among nearby late-type M dwarfs 
 \citep{2000AJ....120.1085G,2004AJ....128..426W,2011AJ....141...97W, 2007AJ....133.2258S}, but both 
metrics decline precipitously for the cooler L and T dwarfs \citep{2002AJ....123.2744B, 2007AJ....133.2258S, 2010AJ....139.1808S} and sources far from the Galactic plane \citep{2006AJ....132.2507W, 2008AJ....135..785W}.  
Similar declines are seen in X-ray emission, but surprisingly not at radio frequencies
\citep{2002ApJ...572..503B, 2005ApJ...626..486B, 2006ApJ...648..629B, 2008A&A...487..317A}.
Assuming that this nonthermal emission arises from magnetic interaction, 
the decline with vertical scale height among M dwarfs is likely an age effect, as angular momentum loss over time results in weakened magnetic dynamos.  However, spin-down timescales exceed 5~Gyr beyond spectral type M5 \citep{2008AJ....135..785W}, and many late-type dwarfs are found to be rapid rotators (P $\approx$ 2-10~hr; \citealt{2003ApJ...583..451M, 2006ApJ...647.1405Z, 2008ApJ...684.1390R}) exhibiting kilogauss magnetic fields at their photospheres \citep{2007ApJ...656.1121R, 2008ApJ...684..644H}.  
Hence, the decline in magnetic emission with spectral type must arise from a different effect.  The favored cause is the decoupling of cooler, increasingly neutral photospheres from magnetic structures, which reduces magnetic stresses and the frequency of magnetic reconnection above the (sub)stellar surface 
(e.g., \citealt{1999A&A...341L..23M, 2002ApJ...571..469M, 2002ApJ...577..433G}).
Deeper reconnection events may continue to power strong flaring bursts observed in a handful of weakly active or inactive late-M and L dwarfs
(e.g., \citealt{1999ApJ...527L.105R, 2003AJ....125..343L, 2007AJ....133.2258S}).

Contrary to these trends, a very rare set of low-temperature ``hyperactive'' dwarfs 
exhibit unusually prodigious and persistent nonthermal emission.
One of the first examples of these to be identified was {\name} (hereafter {\namesh}; \citealt{2002ApJ...564L..89H, 2002ApJ...575..484G}), a high proper motion L5e dwarf which has exhibited
sustained but variable H$\alpha$ emission on no fewer than seven epochs spanning nearly a decade \citep{2002ApJ...564L..89H, 2002ApJ...580L..77H, 2002ApJ...575..484G, 2003AJ....125..343L, 2005A&A...439.1137F, 2006A&A...459..511B, 2008ApJ...689.1295K}. 
With {\lhalbol} $\approx$ $-$4, {\namesh} is as active as a mid-type M dwarf, but is 1--2 orders of magnitude brighter in H$\alpha$
than comparably-classified L dwarfs \citep{2000AJ....120.1085G, 2007AJ....133.2258S}. 
In addition, H$\beta$ and Na~I D lines have also been observed in emission \citep{2005A&A...439.1137F}.  
Since the photosphere of {\namesh} is cool and likely to be highly neutral, the origin of its unexpected emission remains a mystery.  Neither the kinematics nor spectral characteristics of {\namesh} indicate youth, and the absence of mid-infrared excess argues against accretion from a protoplanetary or debris disk \citep{2007ApJ...659..675R}.
Other hyperactive dwarfs, such as 
the L1e dwarf 2MASS~J10224821+5825453 ({\lhalbol} $\approx$ $-$3.5 to $-$2.7; \citealt{2007AJ....133.2258S})
and the T6.5e dwarf 2MASS J1237392+652615 ({\lhalbol} $\approx$ $-$4.6 to $-$4.2; \citealt{2000AJ....120..473B, 2002AJ....123.2744B, 2007ApJ...655..522L}; hereafter 2MASS~J1237+6526) also lack evidence of disk accretion and do not appear to be particularly young.  Alternative mechanisms, such as acoustic heating, unusually strong magnetic fields, and Roche lobe overflow from a substellar companion have been proposed, but none of these scenarios have been validated.

In this article, we report new observations of {\namesh} that reveal both continued nonthermal optical line emission in several neutral atomic species, and the presence of a T dwarf companion at a projected separation of 7~AU.
In Section~2 we describe our combined-light optical and near-infrared spectroscopic observations, the  latter of which yields preliminary evidence for  a brown dwarf companion.  
In Section~3 we describe adaptive optics (AO) imaging and spectroscopic observations that confirm the presence of the companion and allow determination of its separation and classification.   
In Section~4 we analyze the observed and inferred physical properties of the components, the latter based on comparison to evolutionary
and atmospheric models. We also perform a coevality test to examine the reliability of these models.
In Section~5 we discuss how the properties of {\namesh} argue for a magnetic origin to its persistent emission, powered by a strong magnetic field retained by a relatively old and massive cool dwarf.
We summarize our results in Section~6.

\section{Observations: Combined Light Spectroscopy}

\subsection{Magellan/MagE Optical Spectroscopy}

Moderate-resolution optical spectra of {\namesh} were obtained on 2011 March 26 (UT) 
with the Magellan Echellette (MagE; \citealt{2008SPIE.7014E.169M}),
mounted on the 6.5m Landon Clay Telescope at Las Campanas Observatory.  Conditions during the observations were
clear with 0$\farcs$6 seeing.  Two exposures totaling 3000~s were obtained at an average airmass of 1.003 using the 0$\farcs$7 slit aligned with the parallactic angle; this setup provided 3200--10050~{\AA} spectroscopy at a resolution {\ldl} $\approx$ 4000.  We also observed the spectrophotometric flux standard EG~274 \citep{1994PASP..106..566H} on the same night
for flux calibration. ThAr lamps were obtained after each source observation for wavelength calibration, and internal quartz and dome flat field lamps were obtained during the night for pixel response calibration.  Data were reduced using the MASE reduction pipeline \citep{2009PASP..121.1409B}, following standard procedures for order tracing, flat field correction, wavelength calibration (including heliocentric correction), optimal source extraction, order stitching, and flux calibration.

\begin{figure*}
\epsscale{0.9}
\plotone{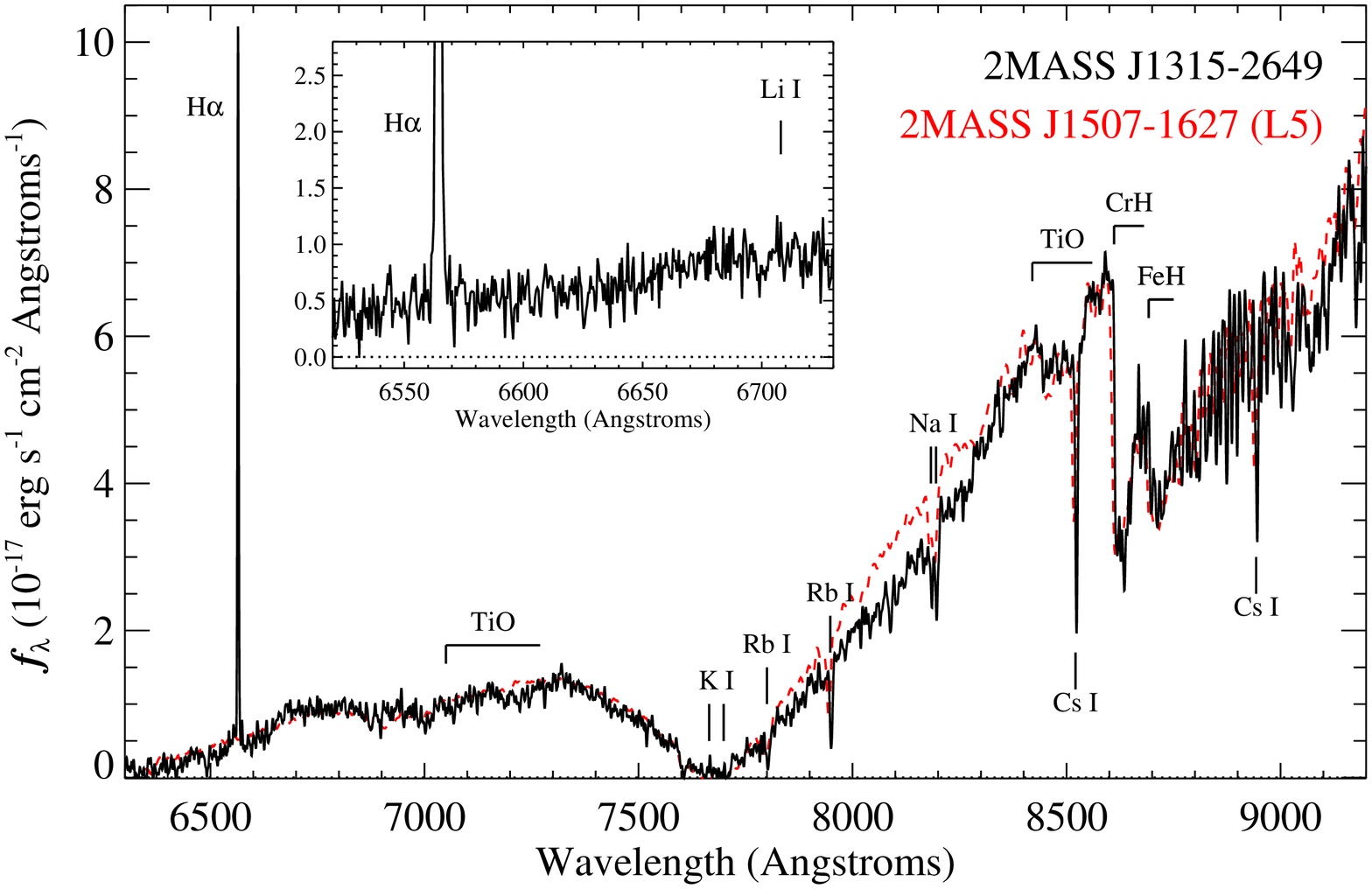}
\caption{MagE red optical spectrum of {\namesh} (black line) compared to the L5 dwarf spectral standard
2MASS~J15074769$-$1627386 (\citealt{2000AJ....119..369R}; data from \citealt{2007ApJ...659..655B}).  Data for {\namesh} are scaled in $f_{\lambda}$ units to its estimated apparent $i$ = 20.16$\pm$0.14 and smoothed to a resolution {\ldl} = 2000; data for 2MASS ~J1507$-$1627 are scaled to align at 8600~{\AA}.   
Note that the apparent discrepancy over 8000--8500~{\AA} is due to uncorrected telluric absorption in the spectrum of {\namesh}.
Absorption features from  K~I, Na~I, Rb~I, Cs~I, TiO, CrH, and FeH are labeled, as is the prominent H$\alpha$ emission.  The inset box shows a close up of the 6500--6750~{\AA} region, highlighting the strong H$\alpha$ line and absence of Li~I absorption.
\label{fig_optspec}}
\end{figure*}

The red portion of the reduced spectrum is shown in Figure~\ref{fig_optspec},
flux-calibrated to an apparent $i$-band magnitude of 20.16$\pm$0.14 as estimated from Two Micron All Sky Survey (2MASS) photometry and a mean $i-J$ = 4.97$\pm$0.13 color for L5 dwarfs \citep{2010AJ....139.1808S}.  
We confirm the characteristic mid-L dwarf features identified in previous studies, including strong FeH and CrH bands; weak TiO absorption (relative to late-M and early-L dwarfs); line absorption from Na~I, Rb~I and Cs~I; and the strongly pressure-broadened 7700~{\AA} K~I doublet.  The overall spectral shape is well-matched
to the L5 dwarf 2MASS J15074769$-$1627386 \citep{2000AJ....119..369R}, consistent with previously reported classifications \citep{2002ApJ...575..484G, 2008ApJ...689.1295K}.   We confirm the absence of  6710~{\AA} Li~I absorption as reported by \citet{2008ApJ...689.1295K} to an equivalent width (EW)  limit of 0.5~{\AA}.  This is well below measured EWs for equivalently-classified L dwarfs (e.g., \citealt{2000AJ....120..447K}).   We also see no evidence of any peculiar spectral features associated with low surface gravities, such as enhanced VO absorption or weakened alkali lines \citep{2006ApJ...639.1120K, 2009AJ....137.3345C}.  
Line center measurements of the alkali lines indicate a heliocentric radial velocity of $-$7$\pm$9~{\kms}.

\begin{figure}
\epsscale{0.9}
\plotone{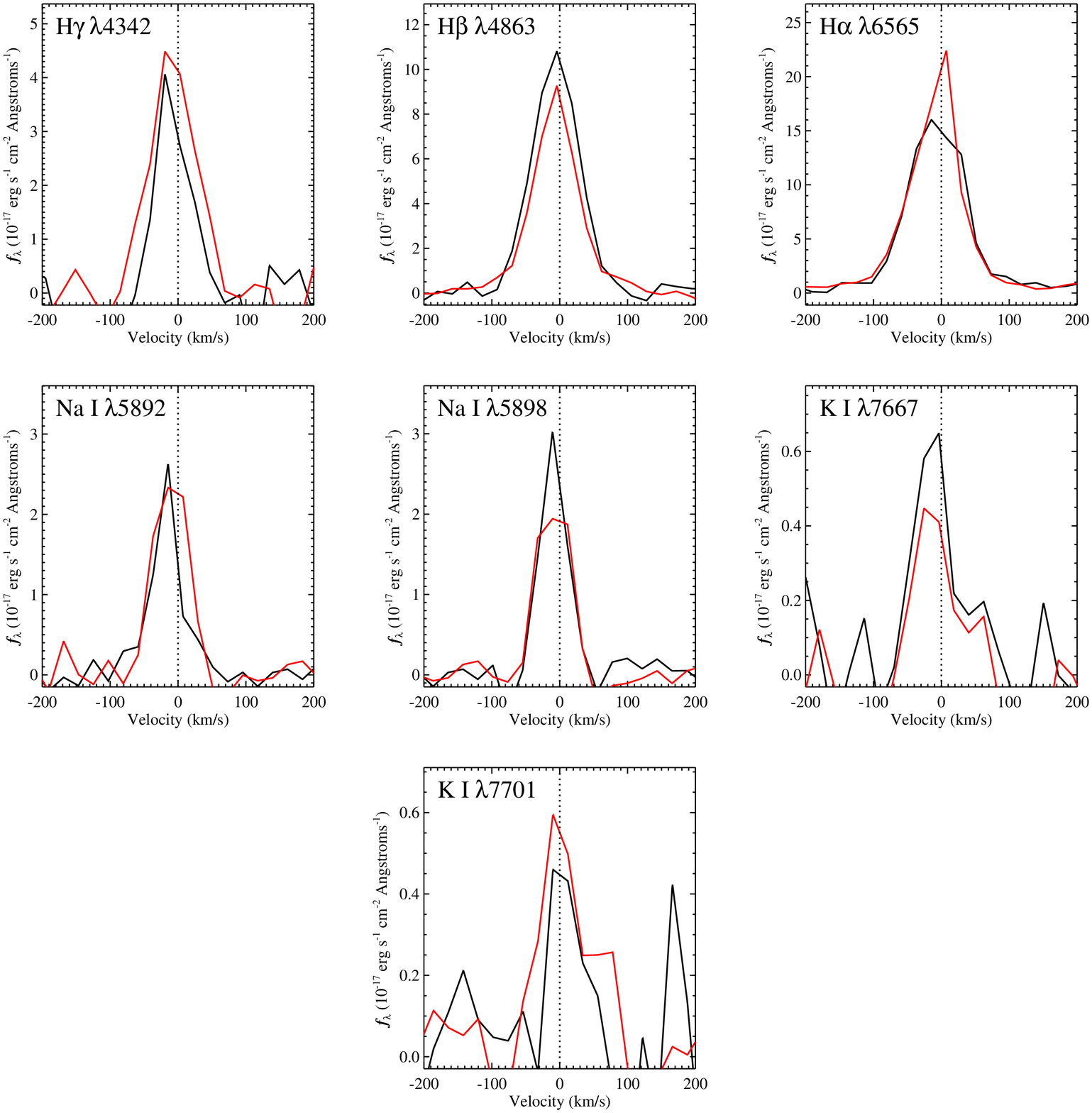}
\caption{Close-up views of emission lines from H$\gamma$ (4342~{\AA}), H$\beta$ (4863~{\AA}), H$\alpha$ (6565~{\AA}), Na~I (5892 and 5898~{\AA}) and K~I (7667 and 7701~{\AA}).  Fluxes are scaled as in Figure~\ref{fig_optspec}, and two separate exposures separated by 30~minutes are shown as black and red lines.
\label{fig_emission}}
\end{figure}

\begin{deluxetable*}{lccc}
\tabletypesize{\footnotesize}
\tablecaption{Optical Line Fluxes for {\name}\label{tab_emission}}
\tablewidth{0pt}
\tablehead{
\colhead{Species} &
\colhead{Equivalent Width} &
\colhead{Line Flux} &
\colhead{$\log_{10}{L_e/L_{bol}}$}  \\
\colhead{} &
\colhead{({\AA})} &
\colhead{(10$^{-17}$ erg~s$^{-1}$~cm$^{-2}$)} &
\colhead{}  \\
}
\startdata
\cline{1-4}
\multicolumn{4}{c}{Absorption} \\
\cline{1-4}
Li~I (6710~{\AA}) & $>-$0.5 & \nodata & \nodata \\
Rb~I (7802~{\AA}) & 4.9$\pm$0.8 & \nodata & \nodata \\
Rb~I (7950~{\AA}) & 5.4$\pm$0.6 & \nodata & \nodata \\
Na~I (8186~{\AA}) & 1.2$\pm$0.3 & \nodata & \nodata \\
Na~I (8197~{\AA}) & 2.5$\pm$0.3 & \nodata & \nodata \\
Cs~I (8523~{\AA}) & 4.2$\pm$0.2 & \nodata & \nodata \\
Cs~I (8946~{\AA}) & 2.9$\pm$0.7 & \nodata & \nodata \\
\cline{1-4}
\multicolumn{4}{c}{Emission} \\
\cline{1-4}
H$\gamma$ (4342~{\AA}) & \nodata\tablenotemark{a} & 4.0$\pm$1.8 & -5.12$\pm$0.20 \\
H$\beta$ (4863~{\AA}) & \nodata\tablenotemark{a} & 13.1$\pm$1.5 & -4.61$\pm$0.08 \\
H$\alpha$ (6565~{\AA}) & $-$58$\pm$4 & 34.9$\pm$1.2 & -4.18$\pm$0.06 \\
Na~I (5892~{\AA}) & \nodata\tablenotemark{a} & 1.6$\pm$0.2 & -5.50$\pm$0.08 \\
Na~I (5898~{\AA}) & \nodata\tablenotemark{a} & 2.1$\pm$0.3 & -5.40$\pm$0.08 \\
K~I (7667~{\AA}) & \nodata\tablenotemark{a} & 0.6$\pm$0.1 & --5.92$\pm$0.10 \\
K~I (7701~{\AA}) & \nodata\tablenotemark{a} & 0.9$\pm$0.2 & --5.79$\pm$0.13 \\
\enddata
\tablenotetext{a}{No continuum available to measure an equivalent width.}
\end{deluxetable*}

The most striking feature in the optical spectrum of {\namesh} is its pronounced H$\alpha$ emission.  We measure an equivalent width of $-$58$\pm$4~{\AA} in the MagE data, in the middle of prior measurements that span $-$24 to $-$160~{\AA}.  We also detect H$\gamma$ (4342~{\AA}) and H$\beta$ (4863~{\AA}) in emission, and confirm the presence of Na~I emission as reported in \citet{2005A&A...439.1137F}.  In addition, we detect weak emission in the cores of the 7700~{\AA} K~I doublets, but no emission from the 5877~{\AA} He~I D3 (EW $<$ 18~{\AA}) or 8500--8660~{\AA} Ca~II triplet lines (EW $<$ 0.3~{\AA}).\footnote{We were unable to quantify the presence or absence of the 3935 and 3970~{\AA} Ca~II  H \& K lines due to an error in the reduction pipeline.  Visual inspection of the spectral images indicates that these lines are not present.} Line profiles of detected emission features (Figure~\ref{fig_emission}) show no appreciable broadening at the 75~{\kms} velocity resolution of MagE, nor do we detect any significant
velocity shift between emission and absorption lines ($\Delta${\vrad} = 4$\pm$16~{\kms}).   
As most of these lines are superimposed on an undetected continuum, we report in Table~\ref{tab_emission} line fluxes and relative line-to-bolometric luminosities, that latter computed using the bolometric correction/spectral type relations of \citet{2010ApJ...722..311L}.   Our measurement of {\lhalbol} = $-$4.18$\pm$0.06 is similar to the first detection made by \citet{2002ApJ...564L..89H}, and is again 1--2 orders of magnitude greater than measured quiescent or flaring fluxes for equivalently classified L dwarfs. 
The Balmer decrement $F_{H\alpha}/F_{H\beta}$ = 2.7$\pm$0.3 is roughly comparable to the mean values for non-flaring M dwarfs \citep{2002AJ....123.3356G}.
H$\alpha$ and H$\beta$ emission contribute 61\% and 23\% of the total measured line flux of $\log_{10}{L_e/L_{bol}}$ $\approx$ $-$4.

\subsection{IRTF/SpeX Near-Infrared Spectroscopy}

Low resolution near-infrared spectra of {\namesh} were obtained on 2009 June 30 (UT) with the 3m NASA Infrared Telescope Facility SpeX spectrograph \citep{2003PASP..115..362R}.  Observing conditions were clear and dry with 0$\farcs$7 seeing at $H$-band.
We used the prism-dispersed mode of SpeX with a 0$\farcs$5 slit (aligned to the parallactic angle) to obtain a continuous 0.7--2.5~$\micron$ spectrum with resolution {\ldl} $\approx 120$.
A total of 8 exposures of 120~s each were obtained in two ABBA dither pairs,
nodding along the slit, at an average airmass of 1.50.
We also observed the A0~V star HD~125438 ($V$ = 7.10) for flux calibration and telluric absorption correction, as well as 
internal flat field and argon arc lamps for pixel response and wavelength calibration.
Data were reduced with the IDL SpeXtool package, version 3.4
\citep{2003PASP..115..389V, 2004PASP..116..362C}, using standard settings; see \citet{2006AJ....131.1007B} for details.

\begin{figure*}
\epsscale{0.9}
\plotone{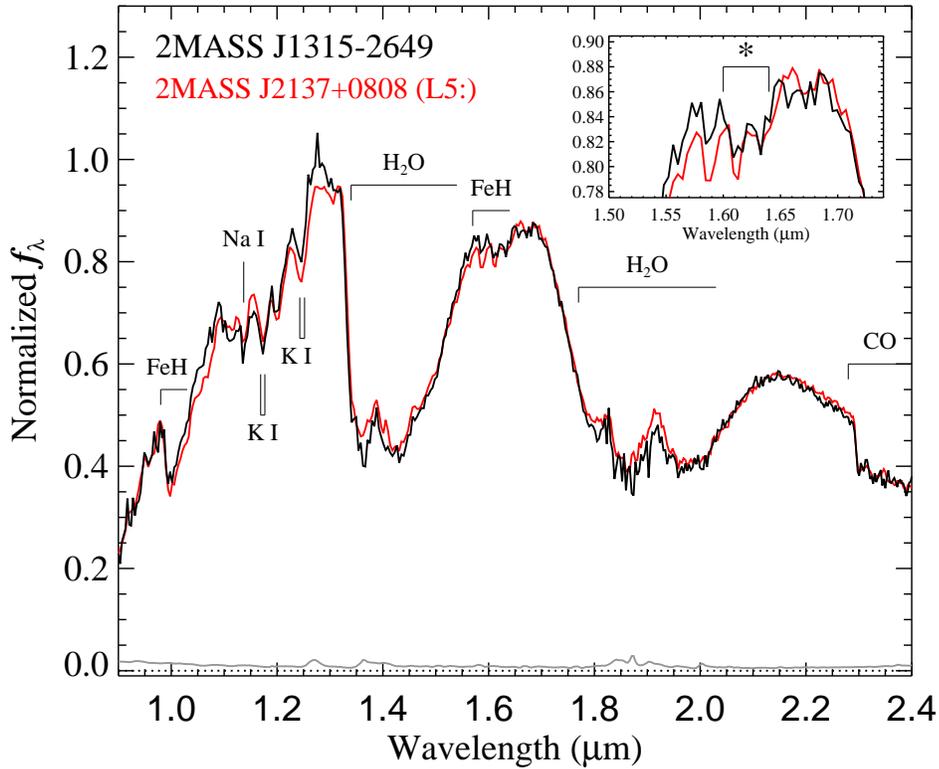}
\caption{SpeX near-infrared spectrum of {\namesh} (black line) compared to the L5 dwarf
2MASS~J2137+0808 (red line; data from A.\ Burgasser et al., in prep.).  Both spectra are normalized at 1.27~$\micron$.  Key spectral features are indicated.  The inset box shows a close-up of the 1.50--1.75~$\micron$ region, highlighting the notch feature that suggests the presence of a T dwarf companion.
\label{fig_nirspec}}
\end{figure*}

The reduced spectrum of {\namesh} is shown in Figure~\ref{fig_nirspec}, compared to the 
L5 dwarf 2MASS~J21373742+0808463 (hereafter 2MASS~J2137+0808; \citealt{2008AJ....136.1290R}).  Its near-infrared spectral morphology is consistent with its L5 optical type, with strong {\wat} and CO absorption bands, FeH absorption at 1.0 and 1.6~$\micron$,
and (unresolved) alkali line absorption in the 1.1--1.3~$\micron$ region.  
The overall spectral shape is again consistent with a normal L5 field dwarf, with no evidence of
an unusual surface gravity, metallicity or cloud content
 (e.g., \citealt{2004ApJ...600.1020M, 2007ApJ...657..511A, 2008ApJ...674..451B, 2008ApJ...686..528L}).

There is, however, a subtle ``notch'' feature present at 1.62~$\micron$ that differs from the spectrum
of 2MASS~J2137+0808 (see inset box of Figure~\ref{fig_nirspec}).
This feature has previously been noted in the combined-light spectra of L dwarf plus T dwarf binaries,
arising from the overlap of FeH absorption in the primary and {\meth} absorption in the secondary (e.g. \citealt{2007AJ....134.1330B, 2008ApJ...681..579B, 2010AJ....140..110G, 2011arXiv1103.1160G}).
To characterize this feature, we performed a spectral template fitting analysis similar to that described in \citet{2008ApJ...681..579B}, using 295 L2--T8 spectral templates from the SpeX Prism Spectral Libraries\footnote{\url{http://www.browndwarfs.org/spexprism}.}.  Fluxes of these templates were scaled to the  $M_{K_s}$/spectral type relation of \citet{2008ApJ...685.1183L}, and all spectra were
interpolated onto a common wavelength scale.  
Restricting potential secondaries to have T spectral types, a total of 17889 binary templates were constructed
and compared to the spectrum of {\namesh} over the wavelength
ranges 0.95--1.35~$\micron$, 1.45--1.80~$\micron$ and 2.00--2.35~$\micron$
using the $\chi^2$ statistic.  

\begin{figure*}
\epsscale{1.1}
\plottwo{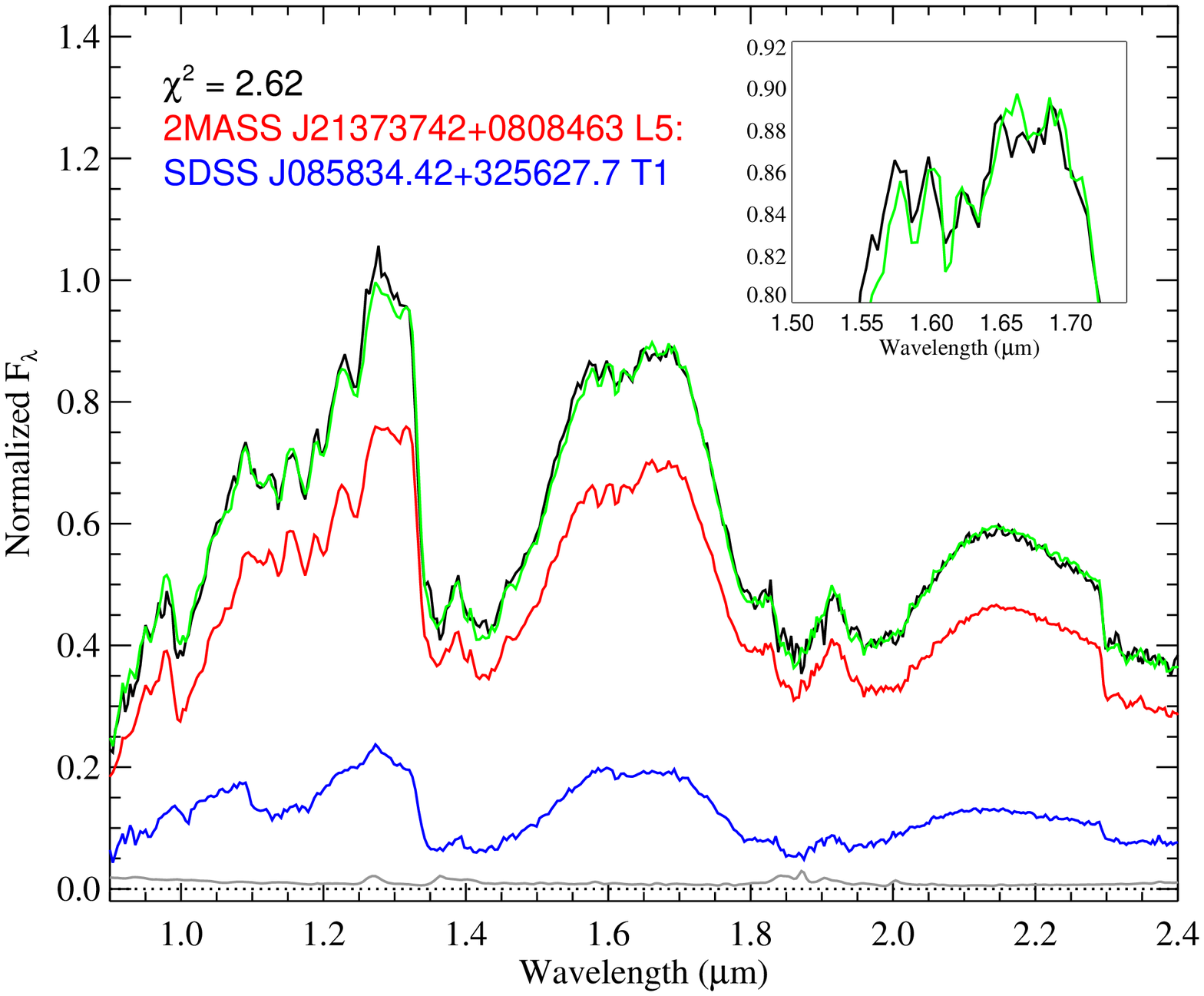}{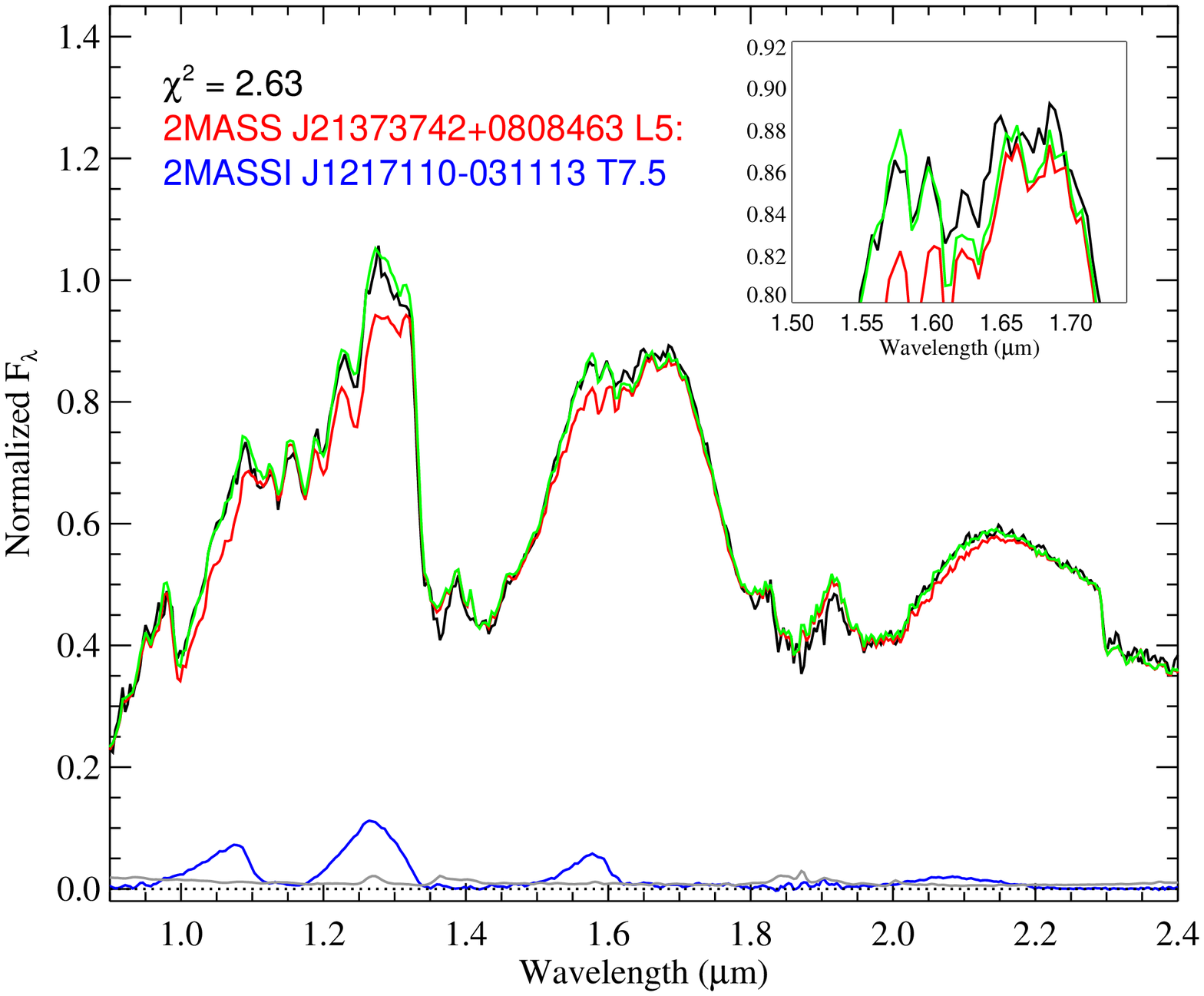} \\
\plottwo{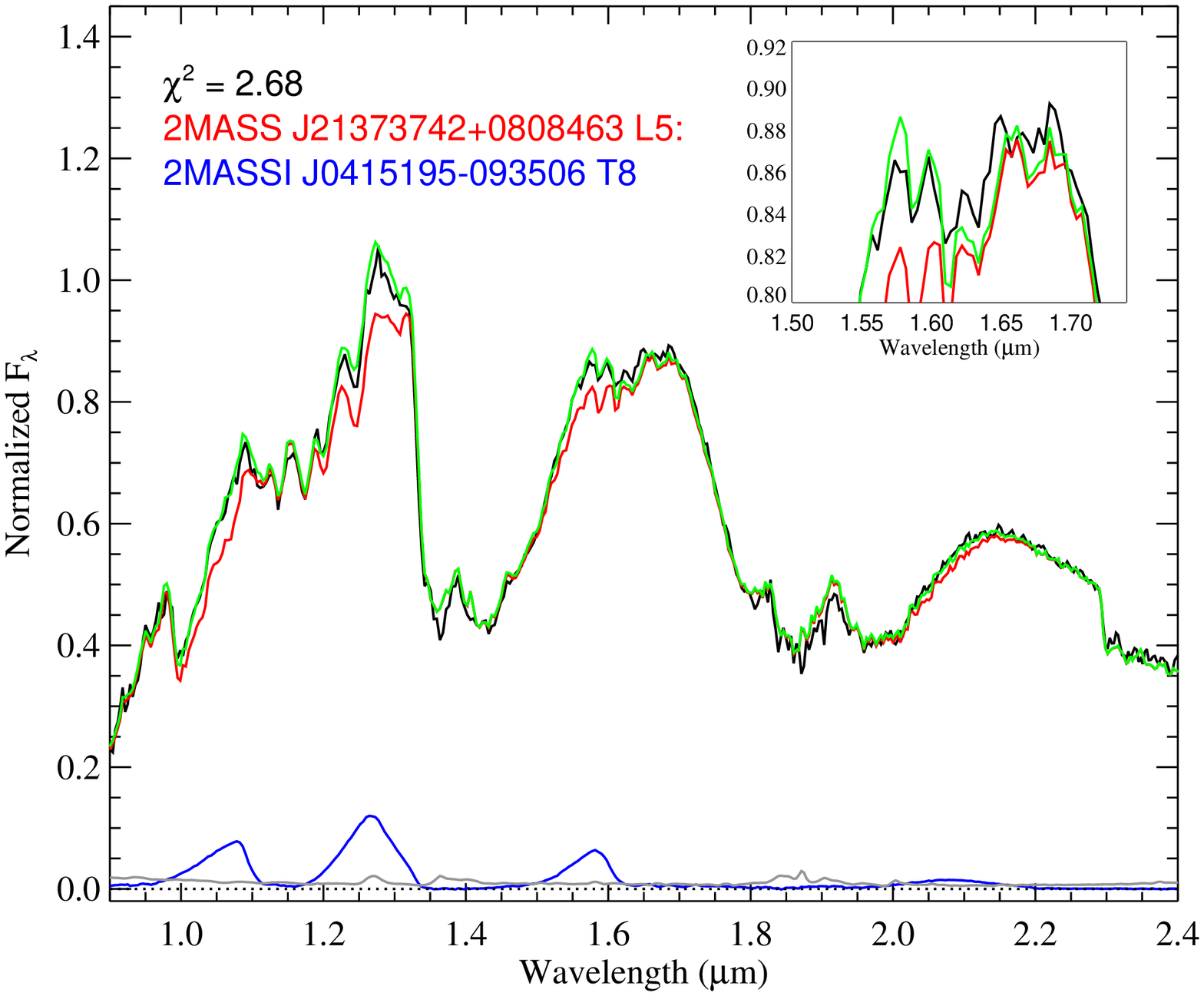}{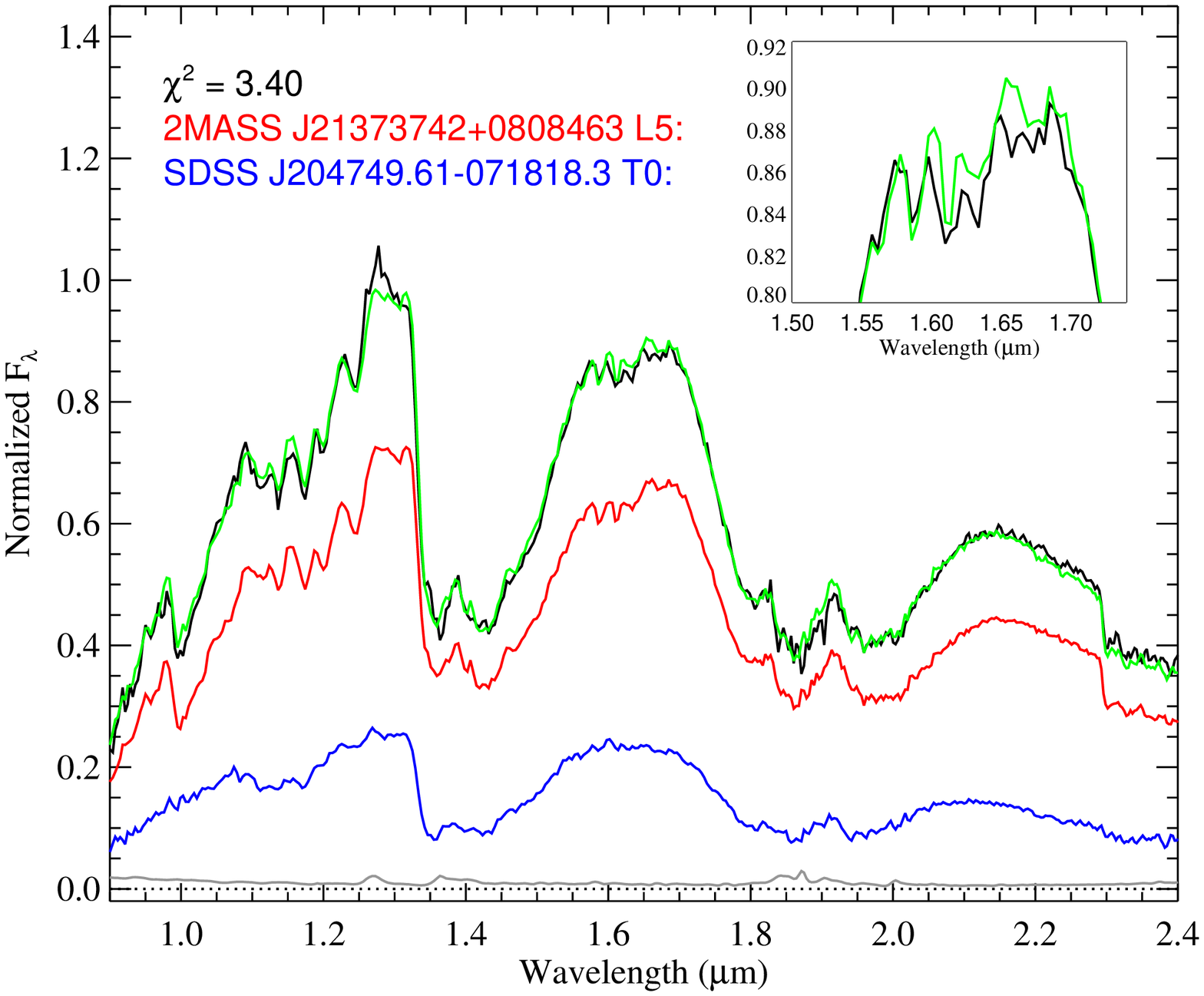}
\caption{The four best-fitting binary templates to SpeX data for {\namesh} (black lines), showing relatively scaled primary (red lines), secondary (blue lines), and combined-light template spectra (green lines).
Component source names and spectral types are listed, along with $\chi^2$ deviations
between template and {\namesh} spectra.
\label{fig_binfit}}
\end{figure*}

Figure~\ref{fig_binfit} shows the four best-fitting binary templates from these comparisons, a
combination of 2MASS~J2137+0808 and either early- or late-type T dwarfs.  The addition of a T-type companion fills in the ``missing'' flux at 1.58~$\micron$, creating the distinct notch feature at 1.62~$\micron$; and at 1.27~$\micron$, producing a somewhat sharper $J$-band flux peak.  Importantly, all of the binary templates shown in Figure~\ref{fig_binfit} provide statistically superior fits to the spectrum of {\namesh} as compared to 2MASS~J2137+0808 alone,
based on the F-test statistic (Eqns.~2--6 in \citealt{2010ApJ...710.1142B}).  However, we cannot precisely constrain the properties of the secondary
from this analysis; the average spectral type of all of the template fits weighted by the F-test probability distribution function (F-PDF)  is T3$\pm$4.  

\section{Resolved Imaging and Spectroscopy}

\subsection{Keck/NIRC2 Near-Infrared Imaging}

To more accurately characterize this putative companion,  high-resolution, near-infrared images of {\namesh} were obtained with the 10m Keck II laser guide star adaptive optics system (LGSAO; \citealt{2006PASP..118..297W, 2006PASP..118..310V}) and facility NIRC2 near-infrared camera.  Observations were conducted on two runs, 2010 March 24 and 2010 May 13 (UT), both with clear skies and fair seeing ($<$1$\arcsec$ and 0$\farcs$5, respectively).
We used the narrow camera   
with image scale 9.963$\pm$0.011~mas~pixel$^{-1}$ \citep{2006ApJ...649..389P} 
covering a 10$\farcs$2$\times$10$\farcs$2 field of view. 
Images were obtained through the $J$, $H$, and $K_s$ filters, using a three-point dither pattern that avoided the noisy, lower left quadrant of the focal plane array.  Exposure times ranged from 30~s with 8~coadds to 120~s with 2 coadds per pointing position, with total integrations of 360~s to 720~s in a given filter.
The sodium LGS provided the wave front reference source for AO correction, while tip-tilt aberrations and quasi-static changes were measured by monitoring the $R$ = 12.8 field star USNO-B1.0 0631-0348160 \citep{2003AJ....125..984M} located $\rho$ = 38$\farcs$6 from {\namesh}. 
Images were reduced using custom IDL\footnote{Interactive Data Language.} scripts, as described in \citet{2010AJ....140..110G}. 

\begin{figure}
\epsscale{0.9}
\centering
\includegraphics[width=0.3\textwidth]{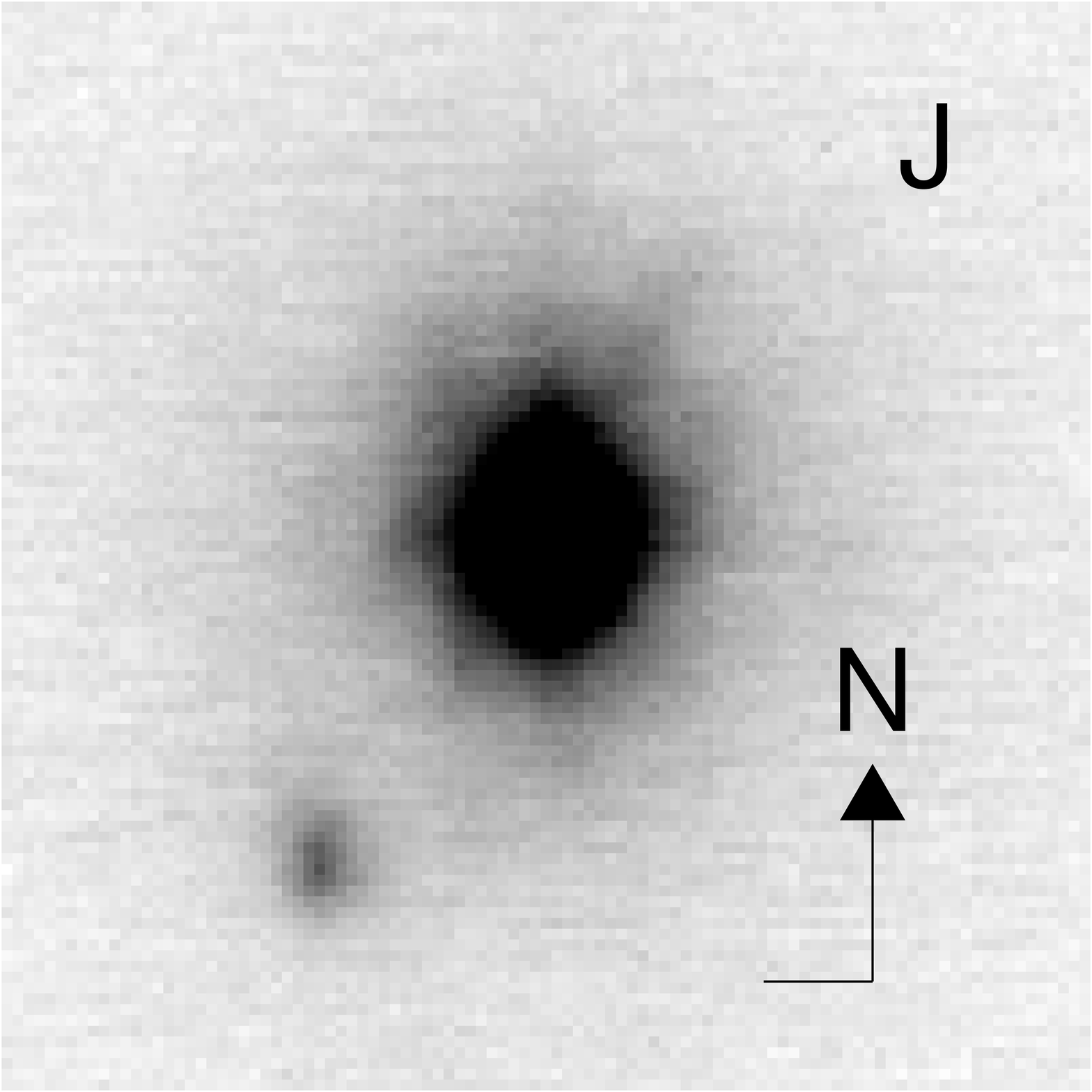}
\includegraphics[width=0.3\textwidth]{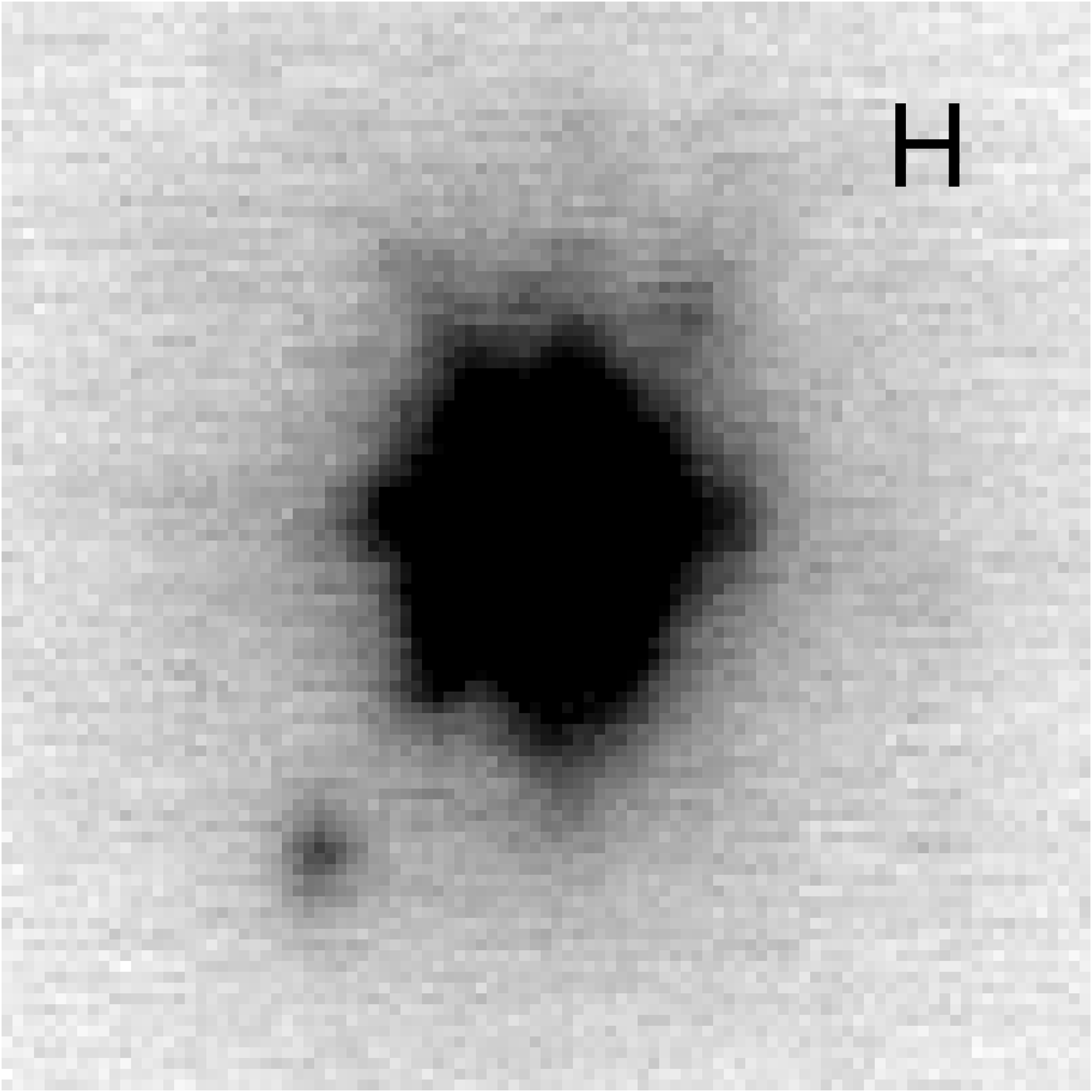}
\includegraphics[width=0.3\textwidth]{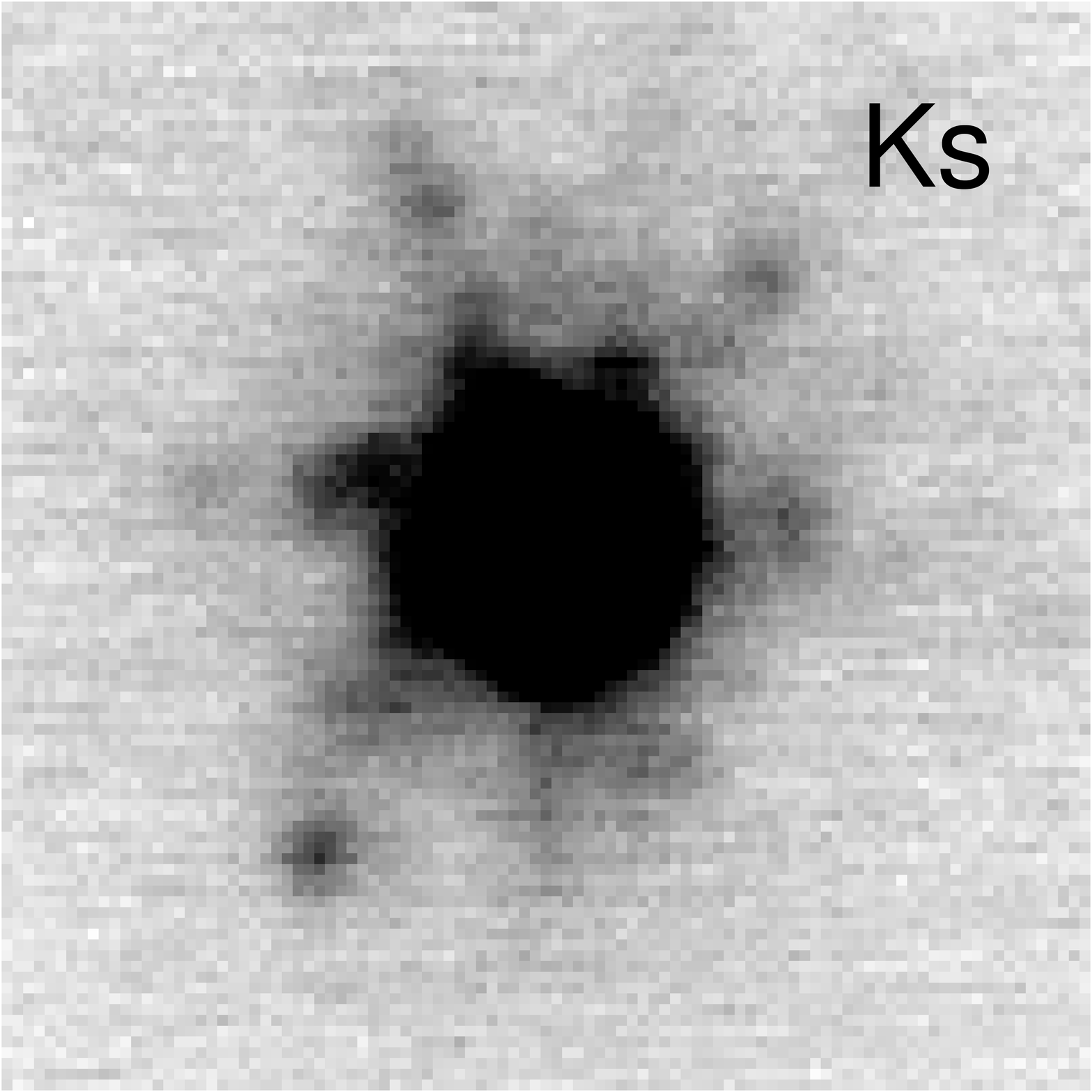}
\caption{1$\arcsec$$\times$1$\arcsec$ NIRC2 $J$, $H$ and $K_s$ images of the {\namesh}AB system.  Images are linearly scaled to optimize visibility of the faint secondary (southeast of the primary), and are oriented with North up and East to the left.   
\label{fig_nirc2}}
\end{figure}

Figure~\ref{fig_nirc2} displays the reduced NIRC2 images from our May observations.
A faint source is clearly present southeast of {\namesh}, and was visible during both imaging epochs. 
Relative astrometry for each epoch was measured on the individual frames 
using a centroiding algorithm, and these 
values were then averaged and multiplied by the pixel scale 
(uncertainties include the standard deviations of the position measurements and 0.1\% pixel scale uncertainty).  
Relative photometry was performed on the coadded mosaics through aperture photometry.  
As the point spread function (PSF) of the brighter component 
contributes significant flux ($\approx$25\%) to the brightness of the 
fainter object, photometry for the latter was extracted from a primary PSF-subtracted image,
constructed by rotating the frame 180$\degr$ about the centroid of the primary 
and differencing.   Systematic errors in the primary subtraction
were estimated by offsetting the rotation axis over a 
5$\times$5-pixel grid centered on the original centroid, subtracting,
and measuring aperture photometry on the secondary.  The
final photometric values for each band and epoch 
were taken as the means and standard deviations of these 25 measurements.  

\begin{deluxetable}{lc}
\tabletypesize{\small}
\tablecaption{NIRC2 and OSIRIS Astrometry and Photometry\label{tab_astrometry}}
\tablewidth{0pt}
\tablehead{
\colhead{Parameter} &
\colhead{Value} \\
}
\startdata
\cline{1-2}
\multicolumn{2}{c}{NIRC2 Epoch 2010 March 24 (UT)} \\
\cline{1-2}
$\Delta{\alpha}\cos{\delta}$ ($\arcsec$) & 185$\pm$6   \\
$\Delta{\delta}$ ($\arcsec$) & $-$277$\pm$3  \\
$\rho$ ($\arcsec$) & 333$\pm$7   \\
$\theta$ ($\deg$) & 143.1$\pm$1.1   \\
$\Delta{J}$ (mag) & 3.12$\pm$0.05 \\
$\Delta{H}$ (mag) & 4.29$\pm$0.14 \\
$\Delta{K_s}$ (mag) & 4.91$\pm$0.18 \\
\cline{1-2}
\multicolumn{2}{c}{NIRC2 Epoch 2010 May 13 (UT)} \\
\cline{1-2}
$\Delta{\alpha}\cos{\delta}$ ($\arcsec$) & 187$\pm$2   \\
$\Delta{\delta}$ ($\arcsec$) & $-$286$\pm$7  \\
$\rho$ ($\arcsec$) & 342$\pm$7   \\
$\theta$ ($\deg$) & 143.9$\pm$0.5   \\
$\Delta{J}$ (mag) & 3.00$\pm$0.03 \\		
$\Delta{H}$ (mag) & 4.59$\pm$0.07 \\		
$\Delta{K_s}$ (mag) & 5.17$\pm$0.12 \\		
\cline{1-2}
\multicolumn{2}{c}{OSIRIS Epoch 2010 May 19 (UT)} \\
\cline{1-2}
$\Delta{\alpha}\cos{\delta}$ ($\arcsec$) & 180$\pm$5   \\
$\Delta{\delta}$ ($\arcsec$) & $-$286$\pm$6  \\
$\rho$ ($\arcsec$) & 338$\pm$4   \\
$\theta$ ($\deg$) & 147.8$\pm$1.2   \\
\enddata
\tablecomments{Angular separation ($\rho$) and position angle ($\theta$) measured from the brighter primary to the fainter secondary.}
\end{deluxetable}

Results are listed in Table~\ref{tab_astrometry}.
Separations in right ascension and declination are consistent between both epochs, 
and (in conjunction with the OSIRIS observations described below) yield a mean separation of 336$\pm$6~mas 
at position angle of $146{\fdg}4{\pm}0{\fdg}5$, measured from primary to secondary.
The mean relative magnitudes are also statistically consistent between epochs (to within 2$\sigma$),
and indicate that the companion is both considerably fainter ($\Delta{J}$ = 3.03$\pm$0.03, $\Delta{K_s}$ = 5.09$\pm$0.10) and bluer in the near-infrared.  To our knowledge, this is the second most extreme near-infrared flux ratio measured for an ultracool dwarf binary to date\footnote{{\namesh} is exceeded at $J$-band only by the young TW Hydrae binary 2MASSW~J1207334-393254AB ($\Delta{J}$ = 7.0$\pm$0.2, $\Delta{K}$ = 4.98$\pm$0.14; \citealt{2004A&A...425L..29C, 2005A&A...438L..25C, 2007ApJ...657.1064M}); and at $K$-band only by the old, widely-separated binary SDSS J141624.08+134826.7AB ($\Delta{J}$ = 4.31$\pm$0.02, $\Delta{K}$ = 6.85$\pm$0.17; \citealt{2010MNRAS.404.1952B, 2010A&A...510L...8S, 2010AJ....139.2448B}).}. We examine the physical association of the companion in Section~4.3; hereafter, we refer to the two sources as {\namesh}A and B.

\subsection{Keck/OSIRIS Near-Infrared Spectroscopy}

Resolved $H$-band spectroscopy of {\namesh}AB was obtained using the Keck II OH-Suppressing InfraRed Integral field Spectrograph (OSIRIS; \citealt{2006SPIE.6269E..42L}) and LGSAO in mostly clear conditions on 2010 May 19 (UT).   We used the 35~mas-scale camera and H$_{bb}$ filter, providing 1.47--1.80~$\micron$ spectroscopy at an average resolution of 3800 and dispersion of 2.1~{\AA}~pixel$^{-1}$ over a 0$\farcs$56$\times$2$\farcs$24 field of view.
The instrument rotator was set at a position angle of $-$45$\degr$ to accommodate both components
in the rectangular field of view.
Six exposures of 600~s each were obtained at an average airmass of 1.52
using a linear dither pattern with steps of 0$\farcs$4 along the long axis, and tip-tilt correction for 
LGSAO operation was again provided by USNO-B1.0 0631-0348160.
These observations were followed by a 600~s exposure of a blank sky frame.
We also obtained three dithered 20~s exposures of the A0~V star HD~107120 ($V$ = 9.90) in natural guide star (NGS) AO mode at an airmass of 1.56 for telluric absorption correction and flux calibration.

\begin{figure}
\epsscale{0.9}
\centering
\plottwo{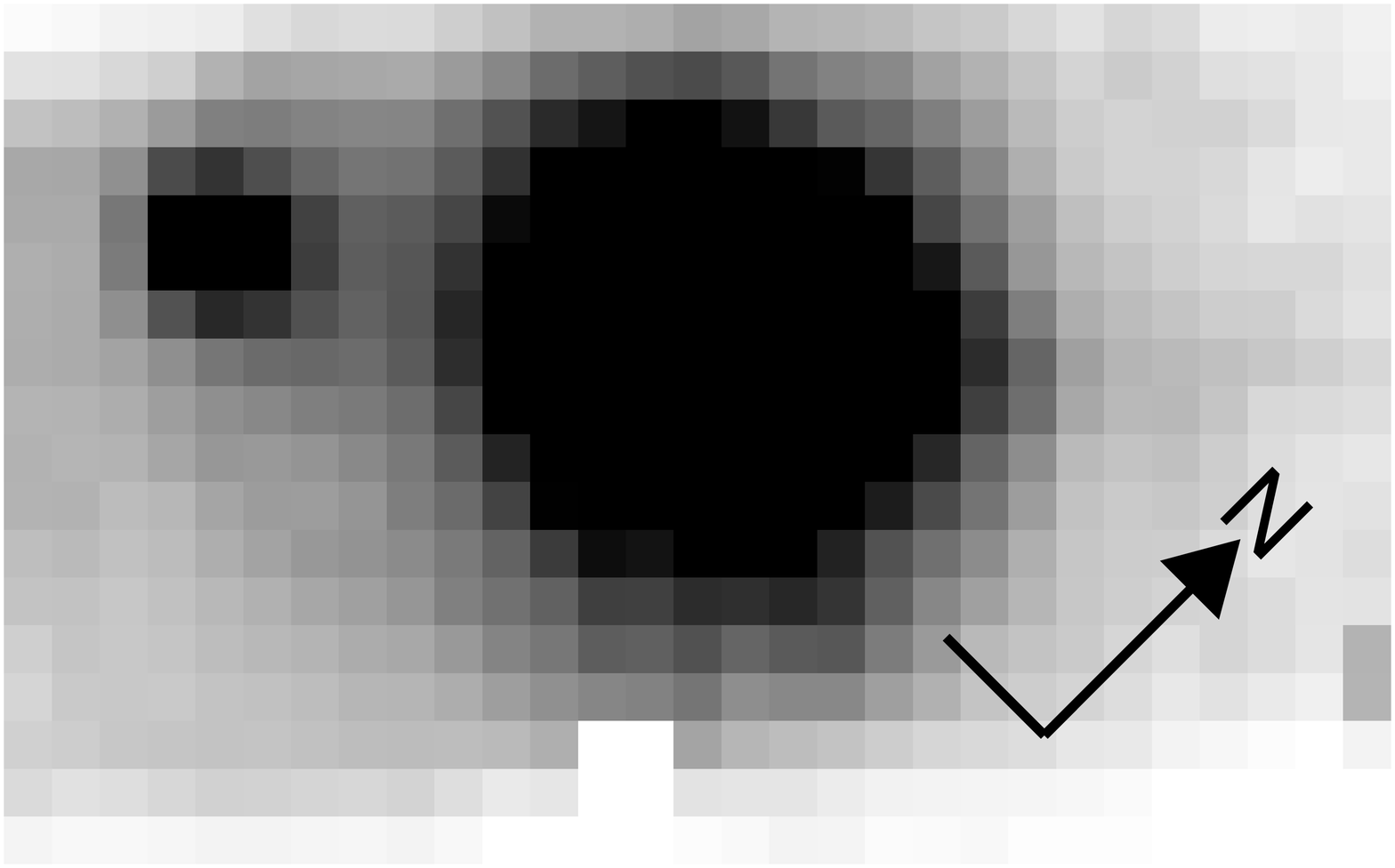}{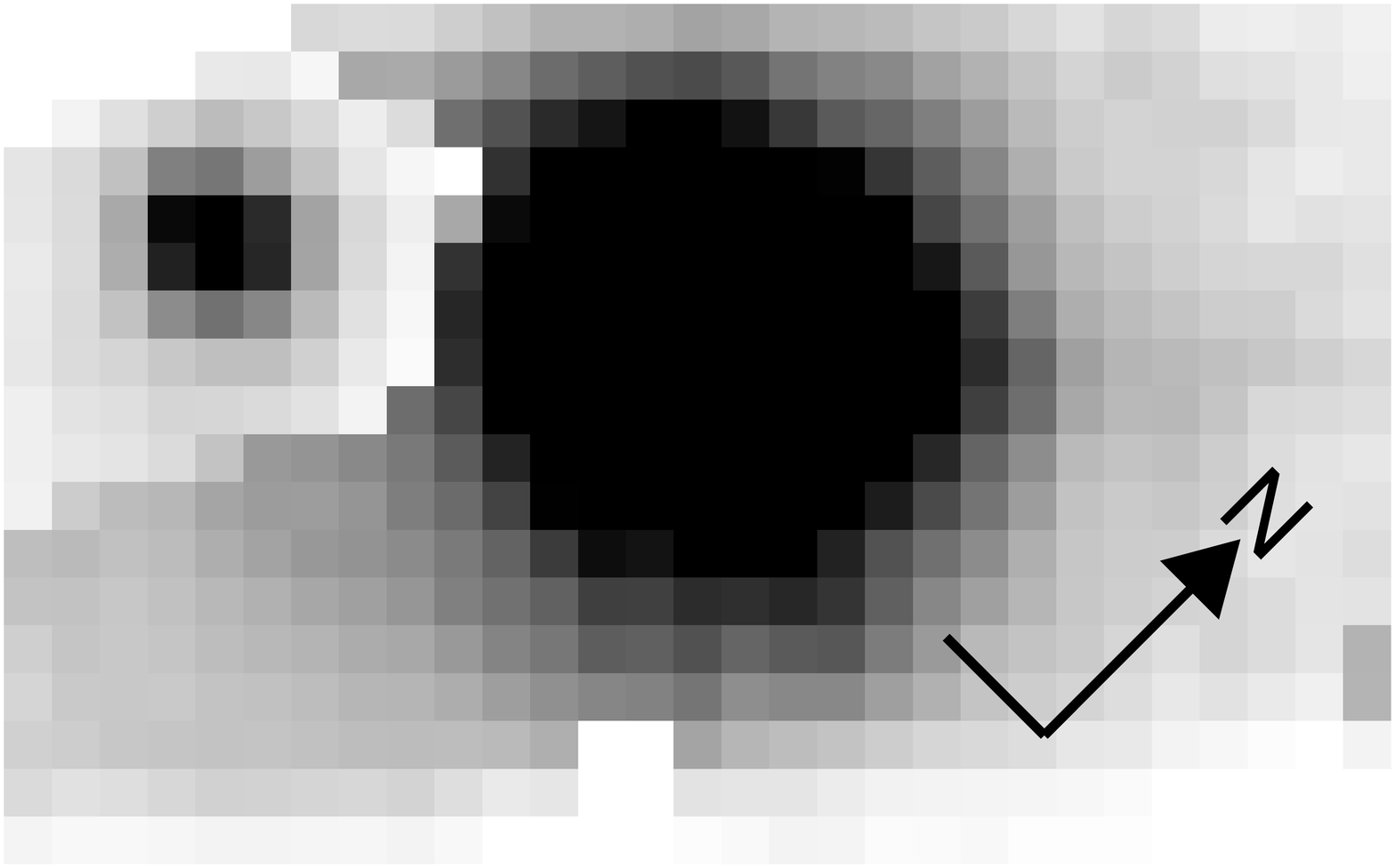}
\caption{Combined mosaics of the OSIRIS data cube for {\namesh}AB, 
spanning a wavelength range of 1.55--1.60~$\micron$. The left panel shows the original data, the right panel shows the result of subtracting a median radial profile of the primary over a position angle range of $\pm$25$\degr$ around the secondary.  The latter image was used to extract the spectrum of the secondary.  The field shown is 1$\farcs$02$\times$0$\farcs$74 and is oriented as indicated by the compass.
\label{fig_osiris_image}}
\end{figure}

Data were reduced with the OSIRIS data reduction pipeline \citep{2004SPIE.5492.1403K}, version 2.3.  We first subtracted the 600~s sky frame from each of the {\namesh} images, and a median-combined dark frame from the calibrator images.  We then used the pipeline to 
adjust bias levels, remove detector artifacts and cosmic rays, extract and wavelength-calibrate the position-dependent spectra (using the most current rectification files as of February 2011), assemble 3D data cubes and correct for dispersion. Spectra for the {\namesh} primary and HD~107120 were extracted directly from the data cube by aperture photometry in each image plane, using a 3-pixel (105~mas) aperture and 10-20~pixel (350--700~mas) sky annulus.  For the faint companion, light contamination from the primary was a concern, so we 
first performed a partial subtraction of the primary's radial brightness profile.  We sampled the profile over two position angle ranges 20--40$\degr$ away from the separation axis, generating a mean profile as a function of wavelength and separation.  We then subtracted this profile $\pm$25$\degr$ about the separation axis.  Figure~\ref{fig_osiris_image} displays mosaics of both the original and subtracted images, illustrating the reduced background achieved around the companion.  We measured aperture photometry for this component in each of the subtracted image planes, 
using a more restricted 1.5~pixel (52~mas) aperture and 3--5~pixel (105-175~mas) sky annulus.
The individual spectra for all three sources were scaled and combined using the {\it xcombspec} routine in SpeXtool \citep{2004PASP..116..362C}.  Flux calibration and telluric correction of the {\namesh}AB spectra were performed using the {\it xtellcor\_general} routine in SpeXtool, assuming a 20~nm Gaussian kernel for
the A0V H~I lines \citep{2003PASP..115..389V}.

\begin{figure}
\epsscale{1.0}
\plotone{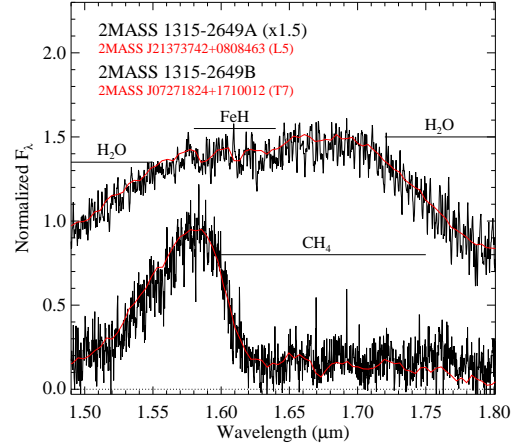}
\caption{Reduced OSIRIS spectra (black lines) of {\namesh}A (top) and B (bottom) over 1.5--1.8~$\micron$,
compared to best-fit SpeX templates (red lines).
Spectra are normalized at 1.58~$\micron$, with data for {\namesh}A and its template scaled by an additional factor of 1.5 for clarity.  FeH, {\wat} and {\meth} absorption bands are labeled.  
\label{fig_osiris_spec}}
\end{figure}

Figure~\ref{fig_osiris_spec} displays the reduced spectra of the two components of {\namesh}.  The spectrum of {\namesh}A has exceptionally high signal-to-noise (S/N $\approx$ 200), and is similar to that of the combined-light SpeX spectrum, with strong {\wat} absorption wings shortward of 1.55~$\micron$ and longward of 1.7~$\micron$, and weak FeH absorption in the 1.57--1.64~$\micron$ region, all indicative of a mid-type L dwarf.  The notch feature, however, is no longer present.   The spectrum of {\namesh}B (S/N $\approx$ 25 at 1.6~$\micron$) is unambiguously that of
a late-type T dwarf, with strong {\meth} absorption at 1.6~$\micron$.

\section{Analysis}

\subsection{Component Spectral Classifications}

Component spectral types were determined by comparing the resolved OSIRIS spectra to  the SpeX Prism Spectral Libraries templates, restricting the template sample to optically-classified
L dwarfs and near-infrared-classified T dwarfs.  
Following a $\chi^2$-fitting procedure similar to that described above over the 1.5--1.8~$\micron$ region, 
we identified the L5 2MASS~J21373742+0808463 \citep{2008AJ....136.1290R}
and the T7 2MASS~J07271824+1710012 \citep{2002ApJ...564..421B}
as the best-fitting templates to {\namesh}A and B, respectively (Figure~\ref{fig_osiris_spec}).   
An F-test weighted average of all templates indicates mean classifications of L3.5$\pm$2.5 and T7$\pm$0.6 for the components.  The large uncertainty for the former is largely due to the broad diversity of near-infrared spectra exhibited by L dwarfs at a given optical spectral type, arising from variations in surface gravity, metallicity and cloud properties
\citep{2007ApJ...657..511A, 2008ApJ...674..451B, 2008ApJ...686..528L, 2010ApJS..190..100K}.  We therefore adopt the optical L5 classification for this component and the near-infrared T7 classification for the secondary.

\subsection{Component Brightnesses, Distances and Luminosities}

Component brightnesses on the MKO\footnote{Mauna Kea Observatory filter system; see \citet{2002PASP..114..180T} and \citet{2002PASP..114..169S}.} system were determined by converting the combined-light 2MASS $JHK_s$ and relative NIRC2 $JHK_s$ magnitudes to MKO $JHK$.  These conversions were computed directly from the SpeX prism spectra of {\namesh}AB (2MASS $\rightarrow$ MKO)
and the best-fit templates in Figure~\ref{fig_osiris_spec} (NIRC2 $\rightarrow$ MKO) using the appropriate filter profiles and a Kurucz model spectrum of Vega (see \citealt{2005ApJ...623.1115C}). 
The resulting combined light and component magnitudes are listed in Tables~\ref{tab_system} and~\ref{tab_components}, respectively.
Both components appear to have relatively normal near-infrared colors for their spectral types 
\citep{2010ApJ...710.1627L}.  

\begin{deluxetable}{lcl}
\tabletypesize{\small}
\tablecaption{Properties of the {\namesh}AB System \label{tab_system}}
\tablewidth{0pt}
\tablehead{
\colhead{Parameter} &
\colhead{Value} &
\colhead{Ref} \\
}
\startdata
Spectral Type & L5 & 1,2 \\
Est.\ Distance (pc) & 19$\pm$3 & 1 \\ 
MKO $J$ (mag)  & 15.07$\pm$0.05 & 1,3 \\
MKO $H$ (mag)  & 14.12$\pm$0.04 & 1,3  \\
MKO $K$ (mag) & 13.45$\pm$0.04 & 1,3 \\
MKO $J-K$ (mag) & 1.63$\pm$0.07 & 1,3 \\
$\mu_{\alpha}\cos{\delta}$ (mas~yr$^{-1}$) & $-$682$\pm$13 & 4 \\
$\mu_{\delta}$ (mas~yr$^{-1}$) & $-$282$\pm$14 & 4 \\
{\vtan} ({\kms})  & 65$\pm$10 & 1,4 \\ 
{\vrad} ({\kms}) & $-$7$\pm$9 & 1 \\ 
U ({\kms}) & $-$38$\pm$8 & 1 \\
V ({\kms}) & $-$41$\pm$9 & 1 \\
W ({\kms}) & $-$13$\pm$5 & 1 \\
$\rho$ (mas)  & 336$\pm$3 & 1 \\
$\rho$ (AU)  & 6.6$\pm$0.9 & 1 \\
$\theta$ ($\degr$)  & 336$\pm$3 & 1 \\
$\Delta{J}$ (mag)  & 3.03$\pm$0.03 & 1 \\ 
$\Delta{H}$ (mag)  & 4.53$\pm$0.06 & 1 \\ 
$\Delta{K_s}$ (mag) & 5.09$\pm$0.10 & 1 \\ 
Age (Gyr) & $\gtrsim$0.8--1.0 & 1 \\
Est.\ Orbit Period\tablenotemark{a} (yr) & 45--60 (15--95) & 1 \\
\enddata
\tablenotetext{a}{First range gives modal values; second range samples the 90\% confidence limits based on Monte Carlo simulation (see footnote~\ref{foot_periodsim}).}
\tablerefs{(1) This paper; (2) \citealt{2002ApJ...575..484G}; (3) 2MASS photometry \citep{2006AJ....131.1163S};
(4) \citealt{2009AJ....137....1F}.}
\end{deluxetable}

\begin{deluxetable*}{lccc}
\tabletypesize{\footnotesize}
\tablecaption{Properties of the {\namesh}AB Components \label{tab_components}}
\tablewidth{0pt}
\tablehead{
\colhead{Parameter} &
\colhead{{\namesh}A} &
\colhead{{\namesh}B} &
\colhead{Difference} \\
}
\startdata
\multicolumn{4}{c}{Observables} \\
\cline{1-4}
NIR SpT & L3.5$\pm$2.5 & T7$\pm$0.6 & \nodata \\
MKO $J$ (mag)  & 15.14$\pm$0.05 & 18.20$\pm$0.06 & \nodata \\ 
MKO $H$ (mag)  & 14.14$\pm$0.03 & 18.66$\pm$0.07 & \nodata \\ 
MKO $K$ (mag) & 13.45$\pm$0.04 & 18.79$\pm$0.11 & \nodata \\ 
MKO $J-K$ (mag) & 1.69$\pm$0.06 & $-$0.59$\pm$0.12 & \nodata \\
Distance (pc) & 18$\pm$4 & 36$\pm$9 & 17$\pm$9 \\ 	
{\lbol} (dex) & -4.19$\pm$0.16 & $-$5.86$\pm$0.16 & \nodata \\
\cline{1-4}
\multicolumn{4}{c}{Spectral Model Fit Parameters} \\
\cline{1-4}
{\teff} (K) & 1760$\pm$70 & 790$\pm$70 & \nodata \\
{\logg} (cm~s$^{-2}$) & $\gtrsim$5.2 & 5.0$\pm$0.5 & \nodata \\
\cline{1-4}
\multicolumn{4}{c}{Evolutionary Models, Age = 1~Gyr} \\
\cline{1-4}
Mass ({\mjup}) & 60$\pm$6 & 16$\pm$3  & 0.26$\pm$0.03\tablenotemark{a} \\
{\teff} (K) & 1720$\pm$150 & 630$\pm$60 & \nodata \\
{\logg} (cm~s$^{-2}$) & 5.26$\pm$0.04 & 4.57$\pm$0.09 & \nodata  \\
Radius ({\rjup}) & 0.90$\pm$0.01 & 1.01$\pm$0.02 &  \nodata \\
\cline{1-4}
\multicolumn{4}{c}{Evolutionary Models, Age = 3~Gyr} \\
\cline{1-4}
Mass ({\mjup}) & 74$\pm$2 & 28$\pm$4 & 0.38$\pm$0.05\tablenotemark{a} \\
{\teff} (K) & 1770$\pm$140 & 670$\pm$70 &  \nodata \\
{\logg} (cm~s$^{-2}$) & 5.41$\pm$0.01 & 4.93$\pm$0.08 & \nodata  \\
Radius ({\rjup}) & 0.84$\pm$0.02 & 0.91$\pm$0.02 &  \nodata \\
\cline{1-4}
\multicolumn{4}{c}{Evolutionary Models, Age = 5~Gyr} \\
\cline{1-4}
Mass ({\mjup}) & 76$\pm$2 & 37$\pm$5 & 0.48$\pm$0.06\tablenotemark{a} \\
{\teff} (K) & 1790$\pm$140 & 690$\pm$70 & \nodata  \\
{\logg} (cm~s$^{-2}$) & 5.44$\pm$0.02 & 5.10$\pm$0.08 &  \nodata \\
Radius ({\rjup}) & 0.83$\pm$0.02 & 0.85$\pm$0.02 &  \nodata \\
\cline{1-4}
\multicolumn{4}{c}{Evolutionary Models, Age = 10~Gyr} \\
\cline{1-4}
Mass ({\mjup}) & 77$\pm$1 & 48$\pm$5 & 0.62$\pm$0.06\tablenotemark{a} \\
{\teff} (K) & 1790$\pm$140 & 720$\pm$70 & \nodata  \\
{\logg} (cm~s$^{-2}$) & 5.44$\pm$0.02 & 5.28$\pm$0.07 & \nodata  \\
Radius ({\rjup}) & 0.83$\pm$0.02 & 0.79$\pm$0.02 & \nodata  \\
\enddata
\tablenotetext{a}{Mass ratio $q \equiv$ M$_2$/M$_1$.}
\end{deluxetable*}

Component distances were computed using the MKO $JHK$ absolute magnitude/spectral type relations of
\citet{2006ApJ...647.1393L}; we considered both the ``bright'' and ``faint'' relations.  Propagating uncertainties in component spectral types and photometry, and scatter in the relations, through Monte Carlo simulation yields consistent distances for each component in all three bands and both relations, with mean values of 18$\pm$4~pc and 35$\pm$9~pc for {\namesh}A and B, respectively.  
These distances are formally consistent with each other, differing at the 1.8$\sigma$ level; the latter has a larger uncertainty due to the larger photometric error for this component (Table~\ref{tab_components}).
The error-weighted mean
distance of 20$\pm$3~pc matches the 21.7~pc estimate of \citet{2007ApJ...659..675R}.
At this distance, the projected separation of the components is 6.6$\pm$0.9~AU, right at the peak of the separation distribution of resolved very low-mass field binaries \citep{2007ApJ...668..492A}.  

Component luminosities were computed from the individual MKO $JHK$ magnitudes using bolometric correction (BC) relations as a function of spectral type, as quantified in \citet{2010ApJ...722..311L}.  Apparent bolometric magnitudes were converted to absolute bolometric magnitudes by adopting a common distance of 20$\pm$3~pc, and luminosities calculated assuming $M_{bol,\sun}$ = 4.74.   Luminosities computed in each of the $JHK$ bands were again mutually consistent, resulting in {\lbol} = $-$4.19$\pm$0.16 and $-$5.86$\pm$0.16 for {\namesh}A and B, respectively.  The luminosity for {\namesh}A is similar to other L4.5--L5.5 field dwarfs
as compiled by \citet{2004AJ....127.3516G} and \citet{2004AJ....127.2948V}, while {\namesh}B is somewhat underluminous for its spectral type.

\subsection{Kinematics and Physical Association}

The similar distances and relative small projected separation of {\namesh}A and B indicates that these sources are 
cospatial; we also find that their space motions are aligned.  The three epochs of NIRC2 and OSIRIS relative astrometry are consistent with each other in both right ascension and declination, and despite the short period between the observations the high proper motion of {\namesh} ($\mu_{\alpha}\cos{\delta}$ = $-$682$\pm$13~mas~yr$^{-1}$, $\mu_{\delta}$ = $-$282$\pm$14~mas~yr$^{-1}$; \citealt{2009AJ....137....1F}) allows us to rule out either component as a (non-moving) background star at the 9$\sigma$ level.  Furthermore, the magnitudes and position angles of the component proper motions are identical to within $\pm$47~mas~yr$^{-1}$ (4.4~{\kms} at 20~pc) and $\pm$4$\degr$.  The radial motions of the two components are also equivalent.  Cross-correlation of the OSIRIS spectra with zero-velocity spectral model templates from \citet[see Section~4.4]{2010arXiv1011.5405A} yield identical velocities to within the uncertainties ($\Delta${\vrad} = 0$\pm$9~{\kms}).  

With a projected tangential velocity of {\vtan} = 70$\pm$18~{\kms}, and adopting the radial velocity from the combined-light optical spectrum above, we find $UVW$ space velocities in the Local Standard of Rest\footnote{Assuming a Solar motion of $U$ = 10~{\kms}, $V$ = 5.25~{\kms} and $W$ = 7.17~{\kms} in the LSR \citep{1998MNRAS.298..387D}, where the directions of $UVW$ follow a right-handed coordinate system.} (LSR) of $U$ = $-$38$\pm$8~{\kms}, $V$ = $-$41$\pm$9~{\kms} and $W$ = $-$13$\pm$5~{\kms}. 
These values lie outside the 1$\sigma$ velocity spheroid for the ``cold'' population of nearby L dwarfs reported by \citet{2010AJ....139.1808S},
but are within
the velocity spheroid of local thick disk stars (e.g., \citealt{2003A&A...398..141S}).  Both the kinematics and spectral properties of {\namesh} are therefore consistent with an older dwarf system in the Galactic disk population.

\subsection{Comparison to Atmospheric and Evolutionary Models}

Assuming that {\namesh}AB comprises a coeval system, insight into the
physical properties of its components can be obtained by joint comparison to atmospheric
and evolutionary models.  The spectral and kinematics analysis above suggests that {\namesh}AB is an older system;
a more quantitative constraint comes from the absence of Li~I absorption
in the combined light optical spectrum, which is dominated by the L5 primary.
The absence of this line sets a minimum mass of $\sim$0.06~{\msun} for this component
\citep{1992ApJ...389L..83R, 1996ApJ...459L..91C}.  Combined with its luminosity, the evolutionary models of
\citet{1997ApJ...491..856B}, \citet{2003A&A...402..701B} and \citet{2008ApJ...689.1327S}
indicate a minimum age for the primary ranging from 0.8 to 1.0~Gyr.  This implies a minimum secondary mass of 0.013~{\msun}, around the deuterium-burning minimum mass limit \citep{2000ARA&A..38..337C,2011ApJ...727...57S}.

\begin{figure*}
\epsscale{1.1}
\centering
\plottwo{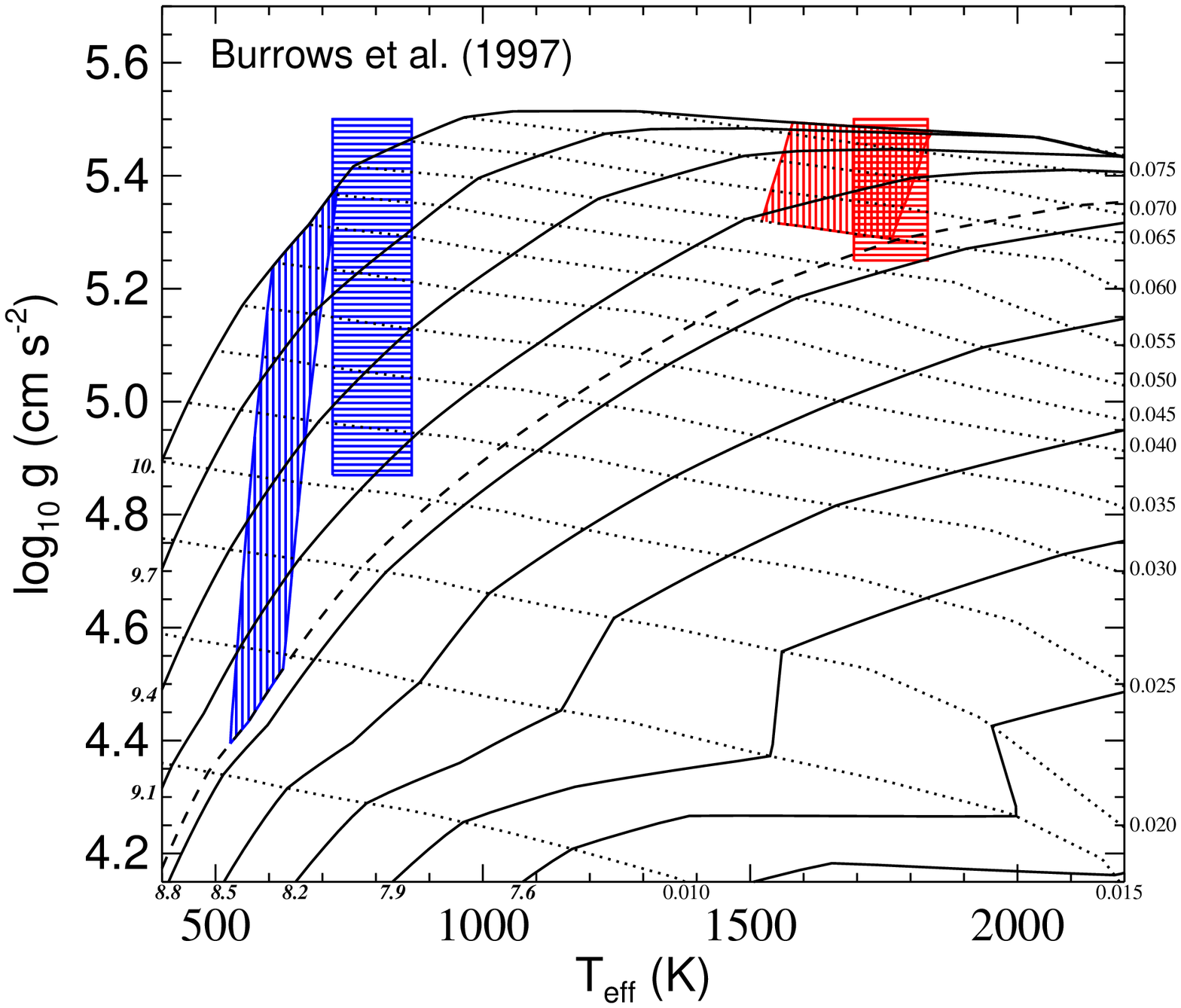}{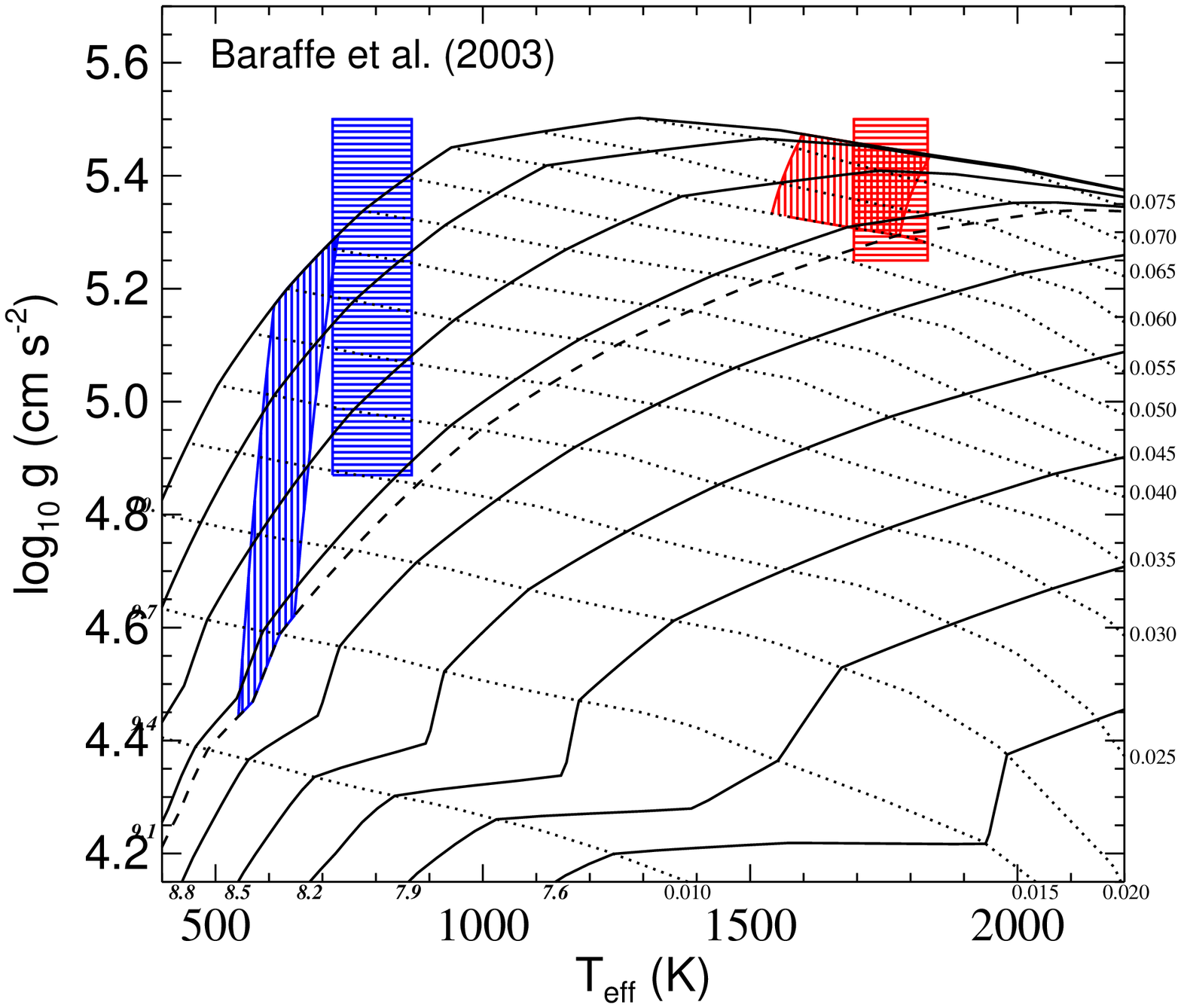} \\
\plottwo{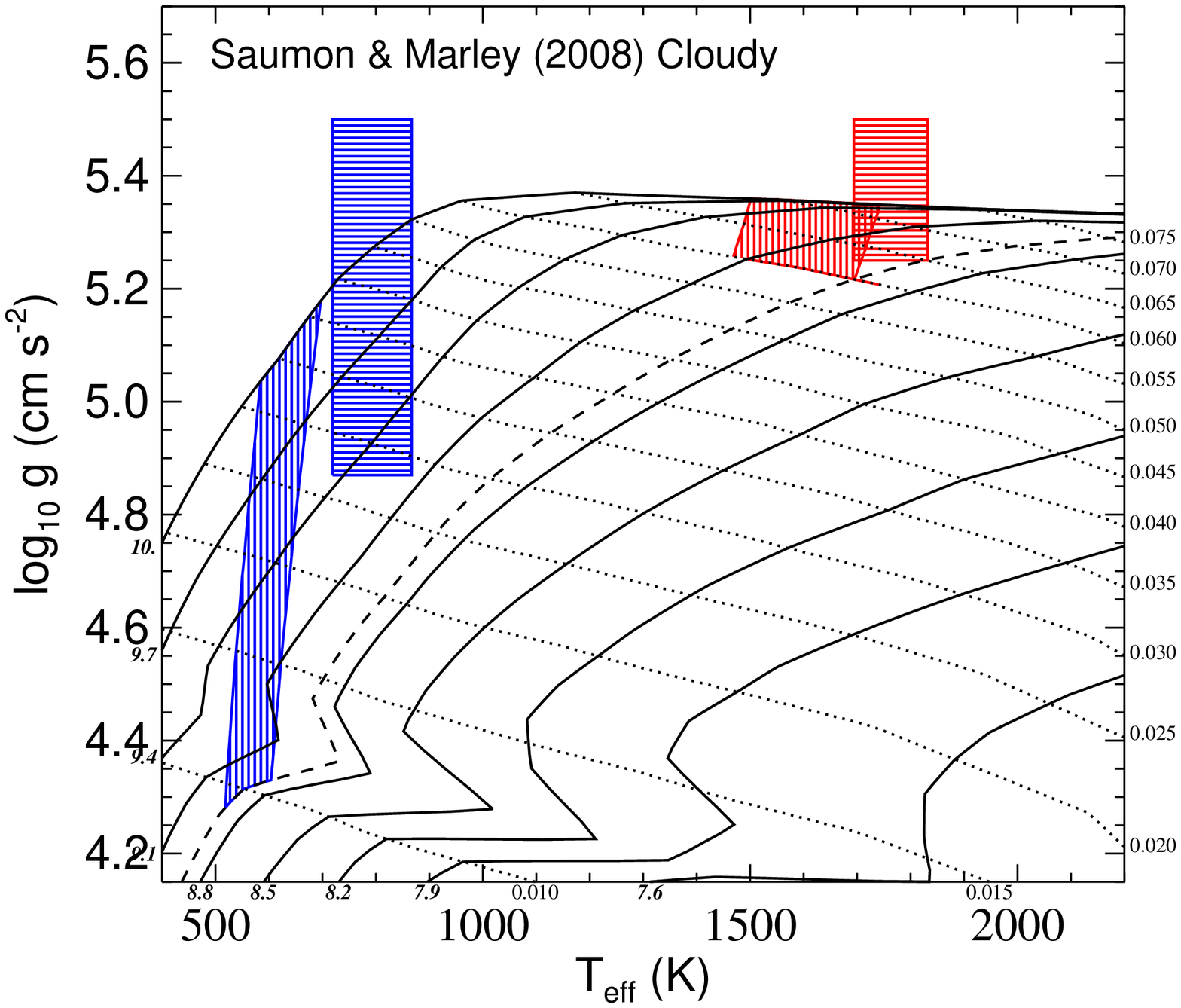}{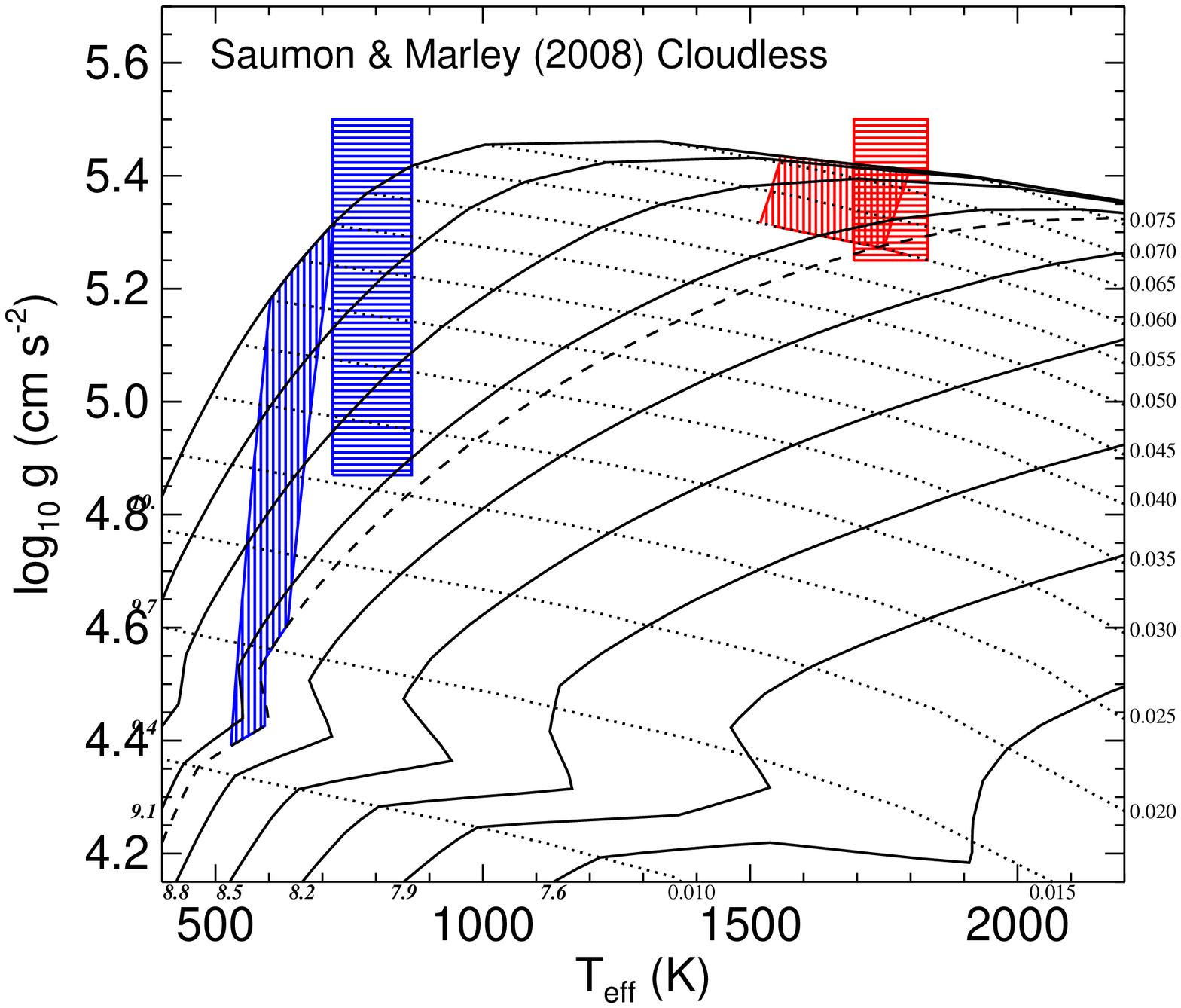}
\caption{Constraints on the {\teff} and {\logg} values of {\namesh}A (at right in red) and B (at left in blue), compared to the evolutionary models of 
(clockwise from upper left) \citet{1997ApJ...491..856B}, \citet{2003A&A...402..701B} and \citet{2008ApJ...689.1327S} (cloudy and cloudless). Solid lines trace isochrones (labeled in $\log_{10}$yr in italics); dotted lines trace isomasses (labeled in solar masses).  The regions constrained by  component  luminosities and the absence of Li~I absorption in the combined light optical spectrum (M$_A$ $\gtrsim$ 0.06~{\msun}) are indicated by vertically hatched regions.  The regions constrained by spectral model fits to the OSIRIS spectra of these components (Figure~\ref{fig_osirisfit}) are indicated by horizontally hatched regions.  The dashed lines trace minimum age isochrones based on the absence of Li~I.
\label{fig_evol}}
\end{figure*}

Figure~\ref{fig_evol} displays the regions in {\teff}/{\logg} space occupied by the components
as constrained by their luminosities, the minimum mass of {\namesh}A, and the evolutionary models listed above.  Table~\ref{tab_components} details specific physical properties for select ages based on the evolutionary models of \citet{2003A&A...402..701B}.  
With no empirical upper limit on the age of this system, the mass of {\namesh}A could be above the hydrogen-burning mass limit (M $\gtrsim$ 0.072~{\msun} for $\tau$ $\gtrsim$ 2~Gyr), while the secondary must be substellar at any age.
The estimated mass ratio of the system, $q \equiv$ M$_2$/M$_1$, is one of the smallest inferred for a very low-mass binary: $\sim$0.3 ($\sim$0.6) for an age of 1~Gyr (10~Gyr).  
This makes {\namesh}AB a unique system, given that $\sim$90\% of 
resolved brown dwarf binaries identified to date have $q > 0.6$ \citep{2007ApJ...668..492A, 2007prpl.conf..427B}.  
Indeed, if low-mass binaries prefer to be in higher mass ratio systems, these values further
support the hypothesis that {\namesh} is quite old.
{\teff} constraints on the components---1570--1930~K for the primary and 570--790~K for the secondary (1$\sigma$ ranges)---are again consistent with comparably classified field dwarfs \citep{2004AJ....127.3516G, 2008ApJ...678.1372C}, with {\namesh}B being somewhat on the cool side for its spectral type.
The lower mass limit for {\namesh}A tightly constrains its surface gravity to {\logg} = 5.22--5.46~{\cmss},
while {\namesh}B has a broader constraint of 4.46--5.35~{\cmss}.

An independent assessment of the component atmospheric parameters was made by fitting the
OSIRIS spectra
to the BT-Settl models of \citet{2010arXiv1011.5405A}. 
These models are based on the PHOENIX code \citep{1999ApJ...525..871H}, and reflect an update
to the original Settl models of \citet{2003IAUS..211..325A} with a microturbulence velocity
field determined from 2D hydrodynamic models \citep{2010A&A...513A..19F} and updated solar abundances
from \citet{2009ARA&A..47..481A}.  
We followed the same fitting procedure described in \citet[see also \citealt{2008ApJ...678.1372C} and \citealt{2009ApJ...706.1114B}]{2010ApJ...725.1405B}, using a set of solar-metallicity models sampling {\teff} = 600--2500~K (100~K steps) and {\logg} = 4.0--5.5~{\cmss} (0.5~{\cmss} steps). 
Model surface fluxes (in $f_{\lambda}$ units) and the OSIRIS spectra were smoothed to a common resolution of {\ldl} = 3500 using a Gaussian kernel, and interpolated onto a common wavelength grid. 
The data were then scaled to the appropriate $H$-band apparent magnitude.
Data and models were compared over the 1.5--1.75~$\micron$ region using a $\chi^2$ statistic, with the degrees of freedom equal to the number of resolution elements sampled.  An optimal scaling factor $C \equiv (R/d)^2$ was computed for each fit to minimize $\chi^2$, where $R$ is the radius of the brown dwarf and $d$ its distance from the Sun \citep{2009ApJ...706.1114B}.
We further constrained our model selection by requiring that the model-inferred distance be within 2$\sigma$ of the estimated spectrophotometric distance of the system (13--26~pc).
Means and uncertainties in the atmospheric parameters were determined using the F-PDF as a weighting factor, as above; we also propagated
sampling uncertainties of 50~K and 0.25~dex for {\teff} and {\logg}, respectively.
Note that these uncertainties quantify {\it experimental} errors; they do not account for {\it systematic} errors that arise from the fidelity of the model fits, as discussed below.

\begin{figure}
\epsscale{0.8}
\plotone{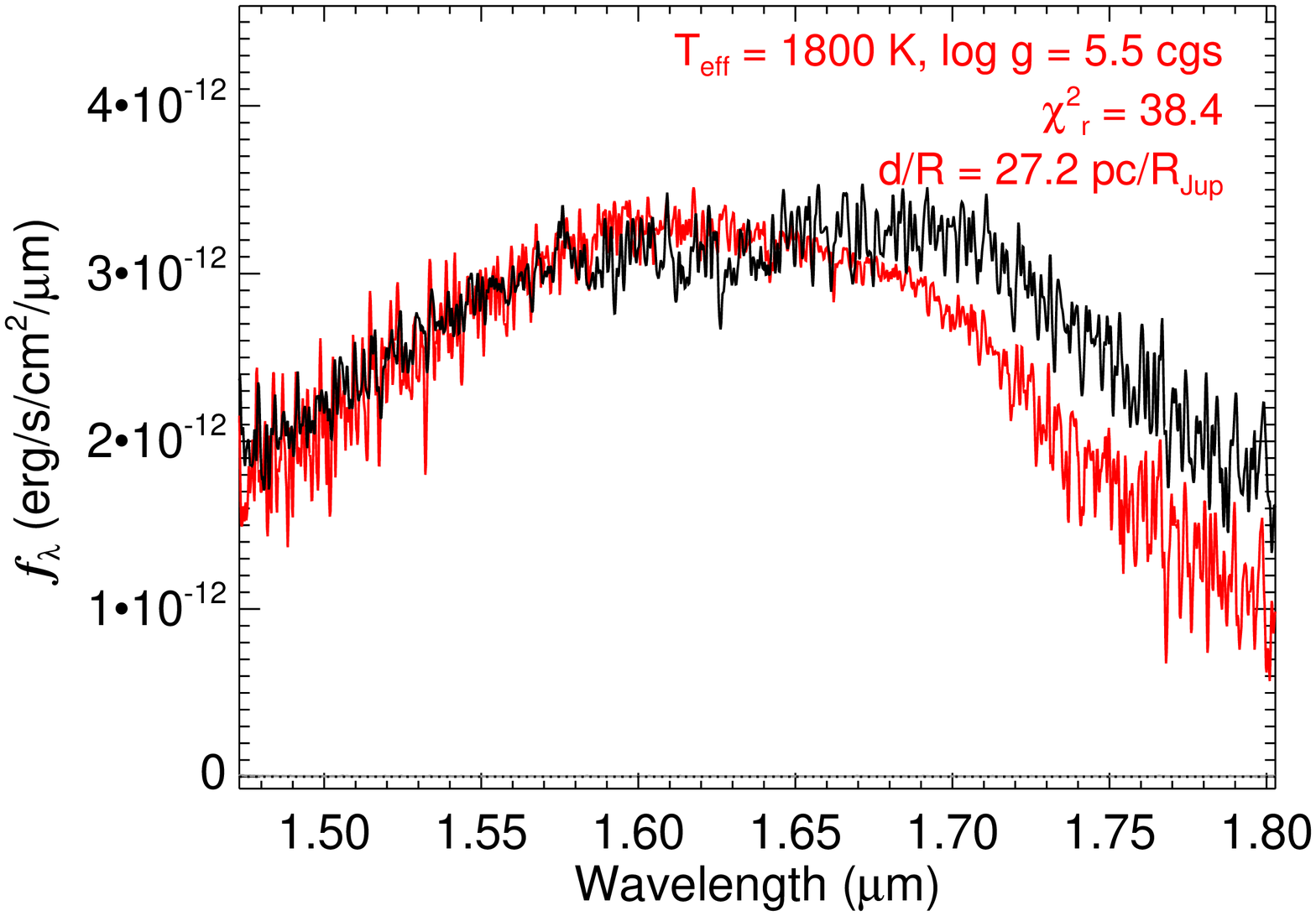}
\plotone{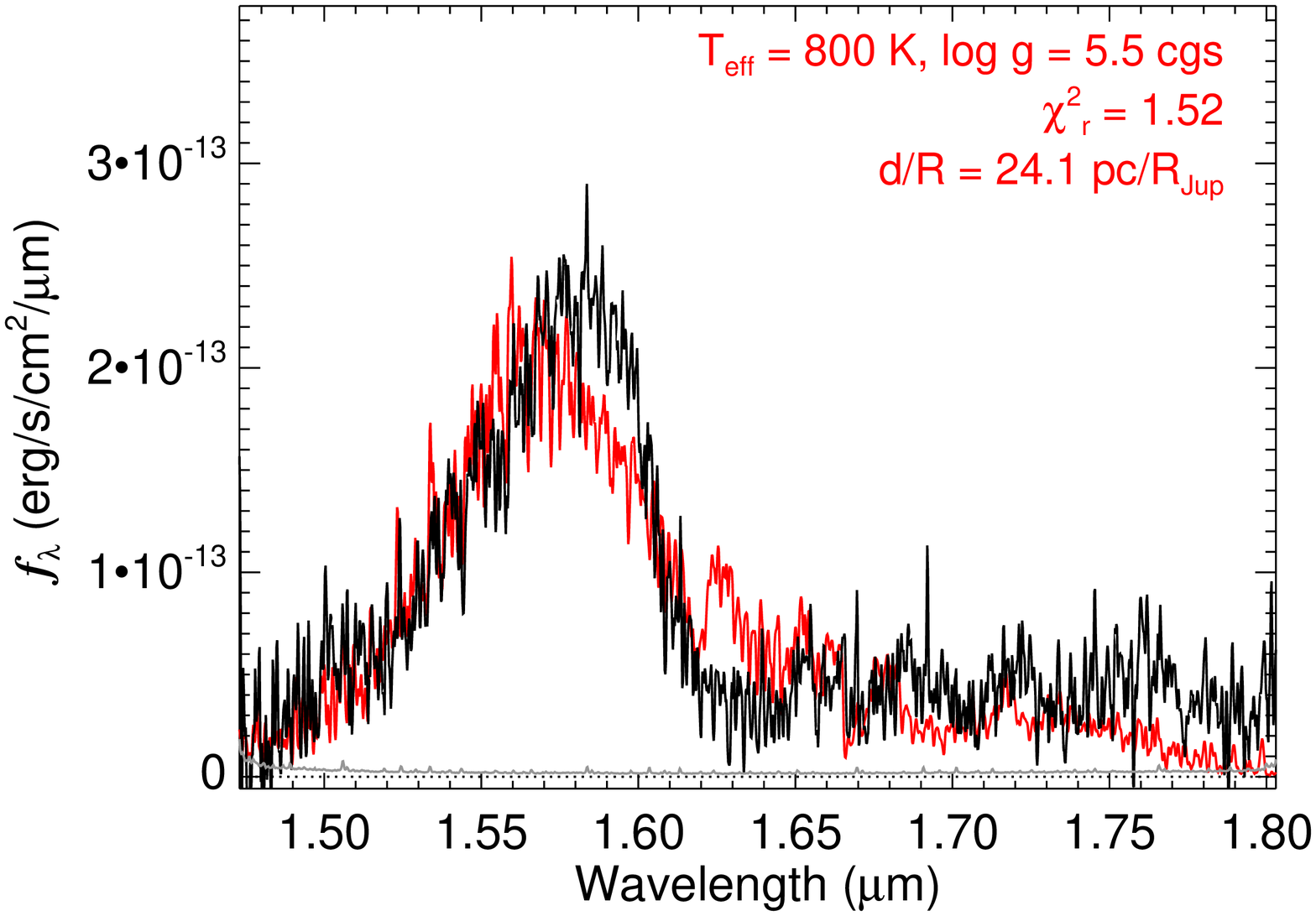}
\caption{Best-fitting BT-Settl spectral models (red lines) for OSIRIS data (black lines) of {\namesh}A (top panel) and {\namesh}B (bottom panel).  The data are scaled to their apparent $H$-band magnitudes, while the models are scaled to minimize the reduced $\chi^2_r$ ($\chi^2$/degrees of freedom).  Model parameters, $\chi^2_r$ and the square roots of the scaling factors (in units of pc/{\rjup}) are indicated in the upper right corners.  Noise spectra are indicated by the grey lines.
\label{fig_osirisfit}}
\end{figure}

Figure~\ref{fig_osirisfit} displays the best-fitting models for the OSIRIS spectra.
For wavelengths shortward of 1.55~$\micron$, the models provide reasonably good fits to the forest
of {\wat} lines present in both component spectra. However, at longer wavelengths we see
deviations in the primary arising from missing FeH opacity and a premature downturn in fluxes
longward of 1.65~$\micron$.  The latter is symptomatic of overly blue spectral energy distributions across the near-infrared range exhibited by the models at these temperatures.  There are also deviations in model fits to the secondary near the 1.58~$\micron$ peak and within the 1.6--1.75~$\micron$
{\meth} absorption system.  Note that, quantitatively, the fit to the secondary is better ($\chi^2_r$ = 1.52) than that to the primary ($\chi^2_r$ = 38.4), but this mainly stems from the high S/N data for the latter; neither fit reproduces the observed spectrum with great fidelity.  Nevertheless, visual inspection confirms that these
are the best fits among the model sample, and the inferred mean parameters---{\teff} = 1760$\pm$70~K and {\logg} $\gtrsim$ 5.2~{\cmss} for the primary and {\teff} = 790$\pm$70~K and {\logg} = 5.2$\pm$0.4~{\cmss} for the secondary---are roughly in line with estimates from the evolutionary model parameters above.

\subsection{A Coevality Test}

With independent determinations of luminosity and {\teff} for both components, we can now examine whether the evolutionary and atmospheric models are
consistent with each other assuming the system is coeval; this is the so-called coevality test (e.g., \citealt{2010ApJ...722..311L}).  Prior studies of brown dwarf binaries have produced mixed results with respect to this test, with a few low-temperature systems showing evidence of $\sim$50--100~K offsets (both high and low) between evolutionary and atmospheric models \citep{2008ApJ...689..436L, 2010ApJ...722..311L, 2009ApJ...699..168D, 2009ApJ...692..729D, 2010ApJ...711.1087K}.  However, spectroscopic {\teff}s have generally been taken from estimates of comparably classified field dwarfs, rather than spectroscopic fits to the binary components themselves\footnote{Note that \citet{2010ApJ...711.1087K} perform atmospheric model fits to resolved component photometry, rather than spectroscopy.}.  Such temperature estimation by proxy could result in systematic biases.  
The single exception is the T1+T6 binary $\epsilon$ Indi BC, for which resolved optical and near-infrared spectroscopy have enabled determinations of individual component luminosities and {\teff}s \citep{2009ApJ...695..788K, 2010A&A...510A..99K}, and comparison of these components on the HR diagram indicate that they are coeval with each other and with their stellar primary \citep{2010ApJ...722..311L}.  

\begin{figure*}
\epsscale{1.1}
\centering
\plottwo{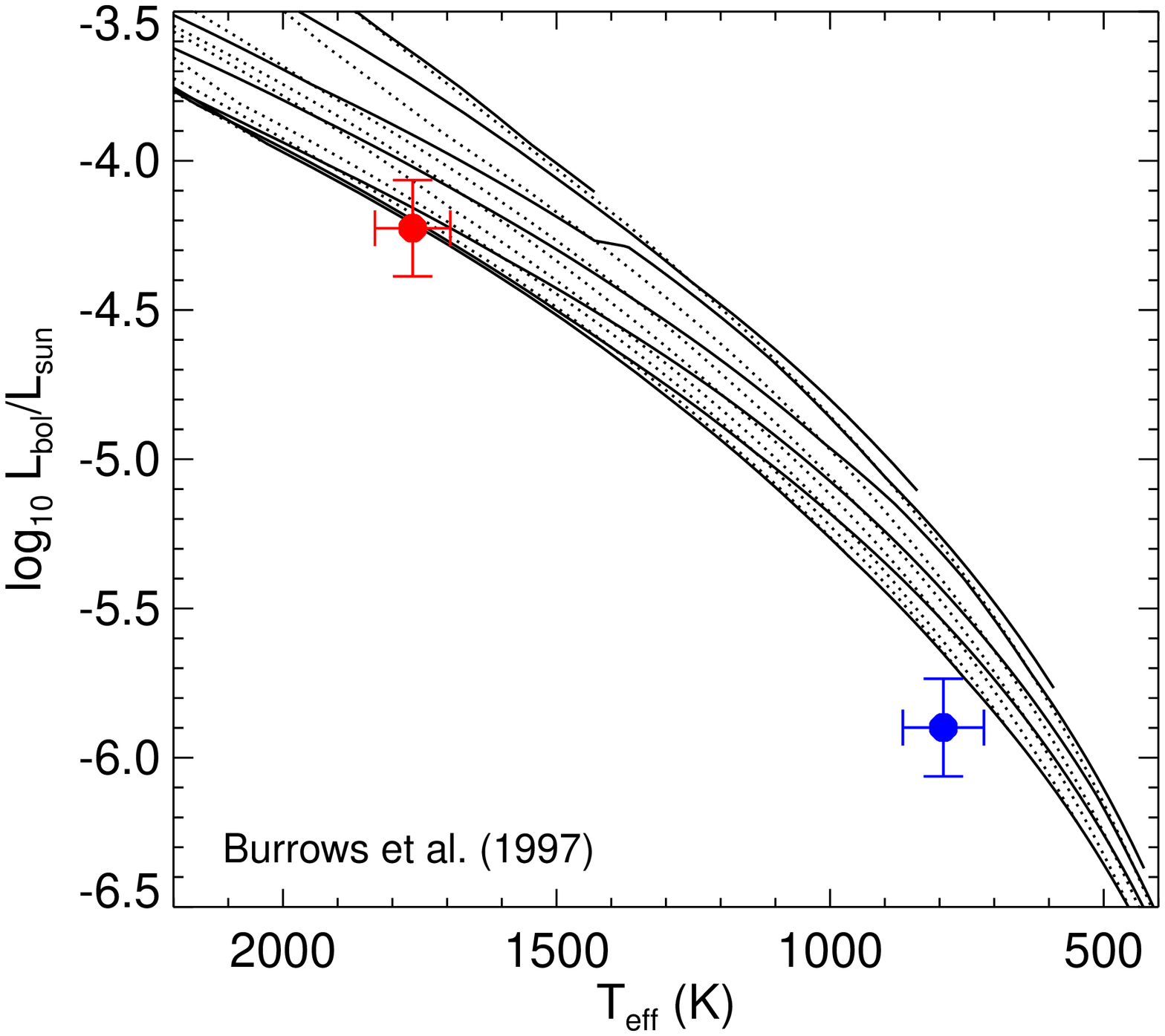}{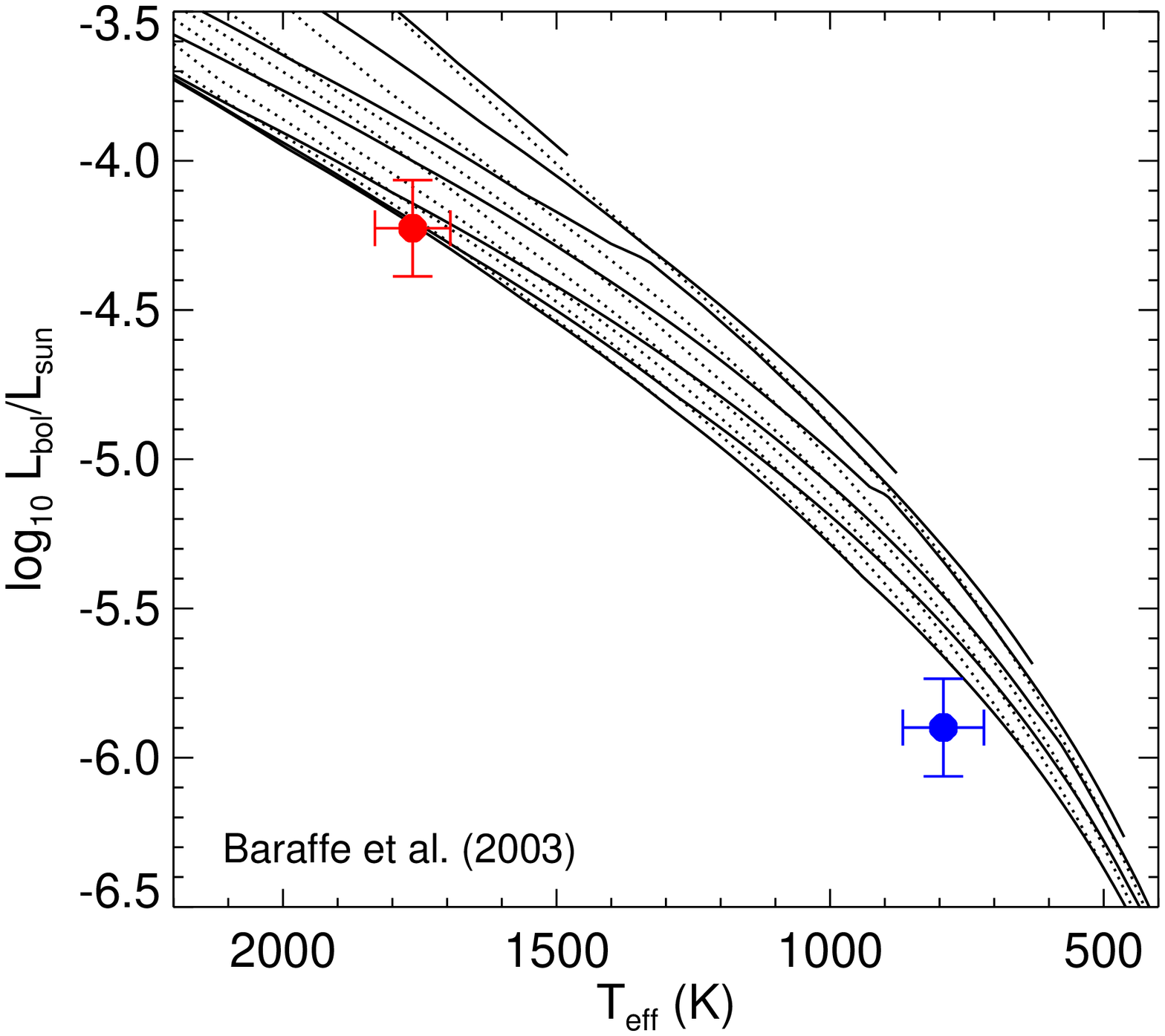} \\
\plottwo{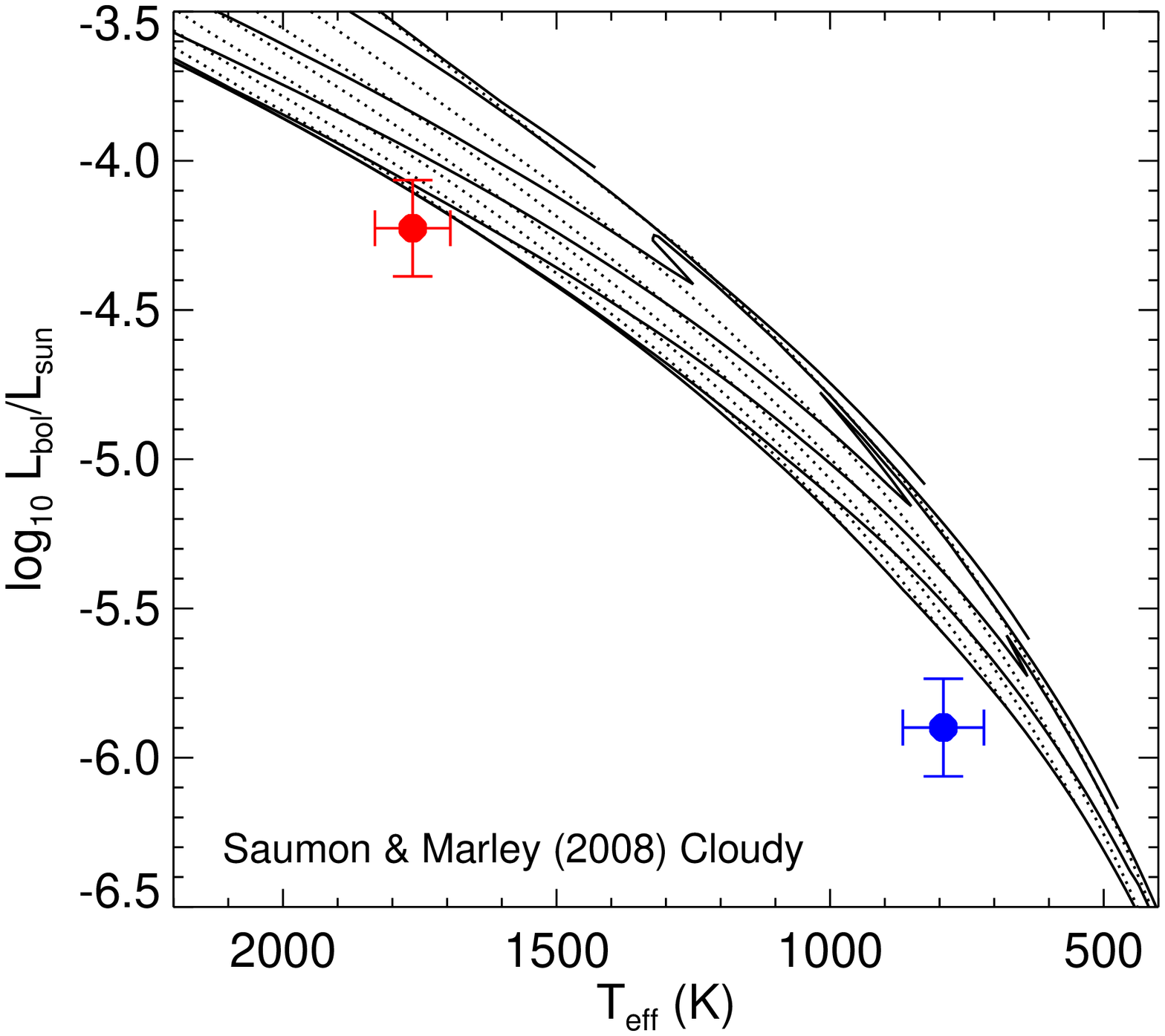}{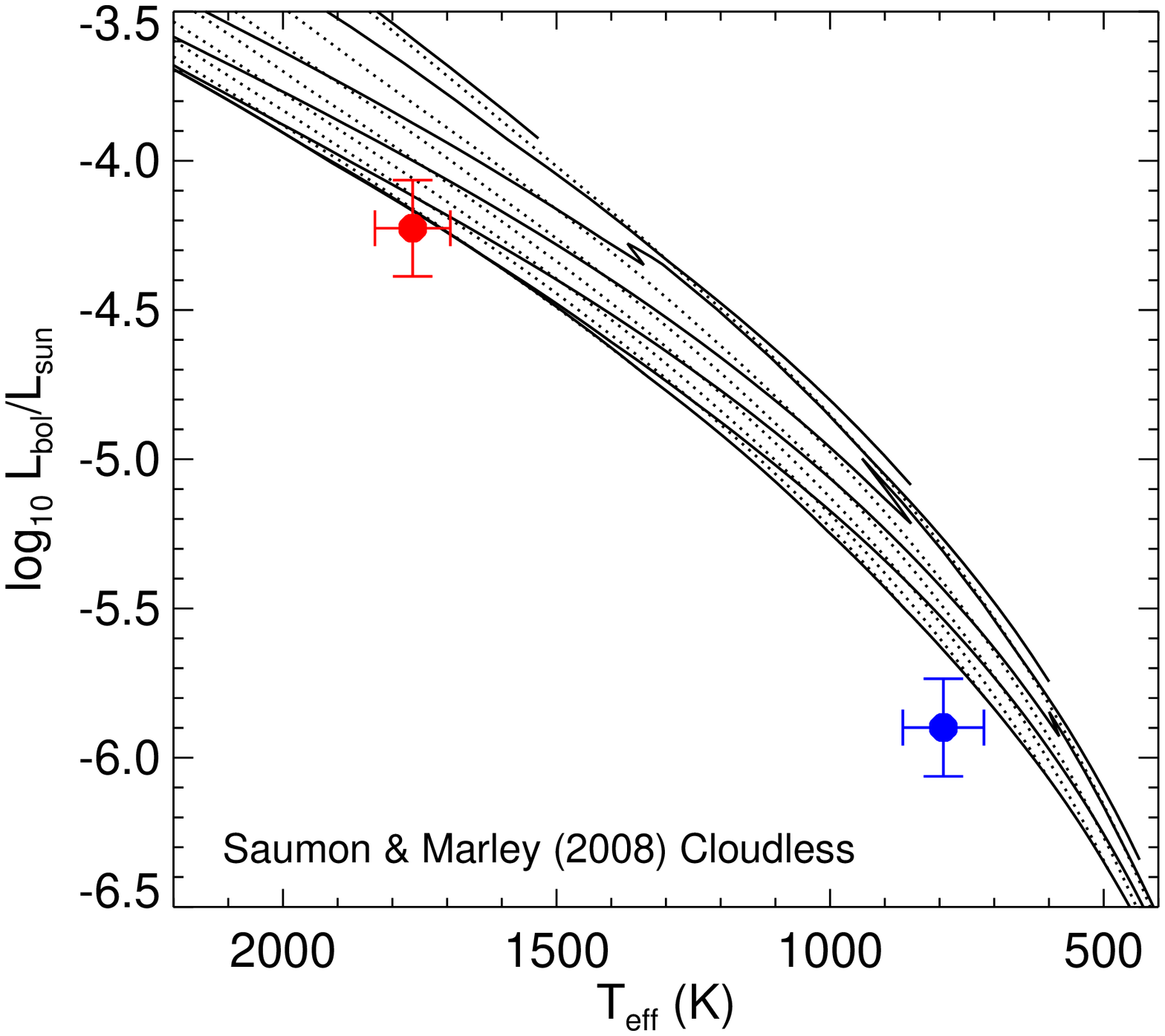}
\caption{Evolutionary tracks in luminosity versus {\teff} (HR diagram) based on the models shown in Figure~\ref{fig_evol}.  The tracks sample the same masses and ages but overlap considerably and are thus not labeled; however, the oldest (most massive) tracks tend to lie along the bottom (left) of the diagrams.  Values for {\namesh}A (top left in red) and {\namesh}B (bottom right in blue),
based on empirical luminosity estimates and spectral model fits, are indicated by the points with error bars.  
\label{fig_hr}}
\end{figure*}

For {\namesh}AB, we find good agreement between atmospheric and evolutionary models for the primary but not for the secondary.   
Figure~\ref{fig_evol} compares the {\teff} and {\logg} constraints from our spectral models fits to those from the evolutionary model comparisons.  The atmospheric {\teff} and {\logg} constraints for the primary overlap reasonably well for all four evolutionary model sets, although there is less agreement for the cloudy models of \citet{2008ApJ...689.1327S}.  Moreover, these values are consistent with the $\sim$0.8-1.0~Gyr minimum age of the system based on the absence of Li~I in the optical spectrum.
For the secondary, overlap in {\teff} and {\logg} regions is not as good, with essentially no overlap for  the \citet{2008ApJ...689.1327S} cloudy models.  This discrepancy can also be seen in Figure~\ref{fig_hr}, which compares the HR diagrams for all four evolutionary models to the luminosities and spectral model fit {\teff}s for the {\namesh} components.  Both sit at the lower envelope of the evolutionary tracks, consistent with older ages; however, {\namesh}B falls off the \citet{2008ApJ...689.1327S} cloudy tracks entirely.

The sense of the deviations between the model comparisons of {\namesh}B is that the spectral model fit {\teff}s are systematically higher than the evolutionary model {\teff}s.  
As we do not have any other independent constraints on the system (e.g., age, metallicity or component masses), we cannot {\it a priori} determine whether this mismatch is specifically attributable to errors in the spectral or evolutionary models.  However, based on the fits shown in Figure~\ref{fig_osirisfit}, we suspect the former given the poor match between the BT-Settl models and spectral data around the 1.6~$\micron$ {\meth} band.  This feature causes persistent problems in spectral model ftis due to incomplete {\meth} opacities at T dwarf temperatures \citep{2006ApJ...647..552S, 2008ApJS..174..504F}.  A decrease of just 100--200~K in the derived secondary {\teff} would bring both components in precise alignment with evolutionary models in both {\teff}/{\logg} and HR diagrams, a shift previously suggested in prior low-mass binary analyses (although not necessarily in the same direction; \citealt{2010ApJ...711.1087K, 2010ApJ...722..311L}). It is also notable that spectral model fits for {\namesh}A are even worse than those for {\namesh}B.  It may be that the alignment of evolutionary and spectral model parameters for this component is merely an example of ``chance agreement''.  In any case, without more accurate fitting of L and T dwarf spectral data, such coevality tests are fundamentally inconclusive about the underlying accuracies of model-derived parameters.

We emphasize that the discrepancies noted here are only at the 1$\sigma$ level, and should be verified through more precise constraints on the component luminosities and {\teff}s.  These can best be accomplished through a parallax distance measurement of the system, as the distance uncertainty dominates the luminosity uncertainty.  In addition, resolved spectroscopy spanning the near-infrared (and possibly optical) range would provide a more robust test of the atmosphere models, and allow us to test different sets of models.  Unfortunately, mass measurements for this widely-separated system are probably not feasible in the near future.   
A probability analysis of the possible orbits for this system\footnote{Period distributions were computed by Monte Carlo simulation, using a method similar to that described in \citet{2007AAS...21110326D}.  Assuming uniform distributions of orbital inclination, ascending node and eccentricity, and using the observed separation as a constraint, we determined probability distributions in eccentric anomaly, semimajor axis and orbit period for various system ages using the masses listed in Table~\ref{tab_components}.\label{foot_periodsim}} predicts likely periods of 45--60~yr (15-95~yr at 90\% confidence).  This rules out a ``fast'' astrometric orbit determination, and the relative orbital radial velocities ($\lesssim$0.5~{\kms}) are comparable to current systematic uncertainties for isolated late-type dwarfs (e.g., \citealt{2010ApJ...723..684B}).  Despite these challenges, the {\namesh} system is an important benchmark for empirical tests of atmospheric and evolutionary models given its proximity, well-separated components, component types and relatively old age.

\section{Why is {\namesh} so Active?}

While the presence of a faint, substellar companion provides useful constraints on the physical properties of the {\namesh} system, it also suggests that a binary interaction could be responsible for the unusual strength and persistence of its nonthermal emission.  In a study of the comparably active T6.5e dwarf 2MASS~J1237+6526, \citet{2000AJ....120..473B}  proposed that accretion of material from a binary companion via Roche Lobe overflow could be a mechanism for sustained emission, as the inverted mass/radius relationship for brown dwarfs allows for sustained mass loss for $q < 0.6$, precisely the mass limit we find for {\namesh}AB.  However, the maximum separation for Roche lobe overflow is only a few Jupiter radii, well below the $\approx$15,000~R$_{Jup}$ projected separation for this pair.  Moreover, the ballistic velocity of material impacting the surface of the primary, $V_B = (2GM/R)^{1/2} \approx 60 (M_J/R_J)^{1/2}$~{\kms} $\approx$ 500~{\kms} (where M$_J$ and R$_J$ are the mass and radius of the primary in Jupiter units) would have been readily detectable from line broadening over a broad range of viewing geometries.  Combined with the absence of infrared excess associated with a circumstellar disk, we rule out accretion from the observed binary companion as the source of line emission.  

The most likely explanation is that {\namesh}A possesses an unusually active chromosphere for its spectral type, perhaps reflecting an unusually strong magnetic field.  
\citet{2000AJ....120.1085G} postulated an inverse relationship between age and activity among late-type M and L dwarfs, finding that stellar-mass (and hence older) L dwarfs were more likely to exhibit H$\alpha$ emission than younger L dwarfs in their sample (note that \citealt{2007AJ....133.2258S} find marginal evidence for the opposite trend).
\citet{2009Natur.457..167C} and \citet{2010A&A...522A..13R} infer a similar mass-dependence
in the strength and persistence of magnetic fields on brown dwarfs based on the total energy flux available for field generation.  In their model, objects above the hydrogen burning mass limit 
retain kilogauss fields for up to 10~Gyr.  
If strong fields correlate with strong chromospheres in L dwarfs (as they do for late-type M dwarfs; \citealt{2010ApJ...710..924R}), then the 
presence of nonthermal emission may align with kinematic, spectroscopic and binary mass ratio evidence that 
{\namesh} is an old system, and {\namesh}A a relatively massive brown dwarf or low-mass star
with a strong magnetic field.
A direct measurement of this component's magnetic field could be obtained through Zeeman line broadening measurements in the 0.99~$\micron$ FeH Wing-Ford band \citep{2007ApJ...656.1121R, 2010ApJ...710..924R, 2010A&A...523A..37S, 2010A&A...523A..58W}. However,  the high spectral resolution required for such a measurement ({\ldl} $>$ 30,000) makes it a challenge for this faint system; indeed, no Zeeman broadening measurements have been reported for an L dwarf to date.
An independent issue is how such a field could generate a persistent chromosphere in the presence of a highly neutral photosphere. Variability in the strength of H$\alpha$ emission from {\namesh} over the past decade may indicate microflaring as a viable source of heating, and a possible indicator of vigorous turbulent field generation \citep{1993SoPh..145..207D, 2006ApJ...638..336D, 2008ApJ...676.1262B}.
Alternately, \citet{2011ApJ...727....4H} have hypothesized that dust grain ionization and electron avalanche (i.e., lightning) could locally increase the photospheric ionization fraction and magnetic field coupling in dusty L dwarfs, although early models appear to favor this mechanism in younger, lower-mass brown dwarfs and exoplanets.  A connection between magnetic activity and ``cloudiness'' has yet to be explored.

For completeness, we note that because of its relatively wide separation, magnetic interaction between  {\namesh}A and B is an unlikely source of emission in this system.
Magnetospheric interactions over scales of up to $\sim$50~R$_*$ have been implicated in outbursts from solar-mass T~Tauri binaries \citep{2006A&A...453..959M, 2008A&A...480..489M}.  However, the estimated $\sim$30~R$_*$ size of a typical L dwarf magnetic field \citep{2009ApJ...699L.148S}\footnote{The field size is take to be the Chapman-Ferraro radius, where the magnetic field pressure balances ram pressure from the interstellar medium.  This size scales as a weak function of the magnetic field strength, with R$_{CF}$/R$_{*}$ $\propto$ B$^{1/3}$.} is still several orders of magnitude smaller than the  projected separation of the {\namesh} binary.  In addition, the possibility that the observed nonthermal emission arises from the secondary, rather than the primary, is unlikely.  This scenario would imply a relative emission luminosity of {\lhalbol} $\approx$ $-$2.5 for the secondary, an amplitude seen only during exceptionally large flare bursts from M and L dwarfs (e.g., \citealt{2007AJ....133.2258S, 2010AJ....140.1402H}).  While other interactions could be postulated (e.g., magnetic wind interactions, Kozai-like perturbations of an unseen third component), these scenarios are far more complex than the simple hypothesis of enhanced chromospheric emission powered by a strong magnetic field around a relatively old and massive cool dwarf.   In other words, while the presence of a low-mass T dwarf secondary in this system is certainly intriguing, it appears to play no role in the observed nonthermal emission.

\section{Summary}

We have identified a T7 brown dwarf companion to the unusually active L5e dwarf {\namesh}, a source that continues to exhibit strong H~I and alkali line emission despite its late spectral type.  
Resolved imaging and spectroscopy confirm the pair to be co-moving and co-spatial, 
and evolutionary models indicate an unusually low mass ratio as compared to other low-mass field binaries. 
The spectral and kinematic properties of both components confirm prior indications that this system is relatively old ($\tau$ $\gtrsim$ 0.8--1.0~Gyr), and likely a member of the old Galactic disk population.
Joint comparison to atmospheric and evolutionary models (a coevality test) also supports an older age for this system, but reveals discrepancies in the case of the secondary; we suspect these are due to continued shortcomings in the modeling cool brown dwarf spectra.
The age and separation of the system, coupled with the narrow emission lines and absence of mid-infrared excess, rule out accretion from or magnetic interaction with the T dwarf secondary as the emission source.  
Rather, we attribute it to an unusually strong magnetic field as predicted by 
energy flux scaling arguments
for relatively old and massive low-mass dwarfs.
A direct test of this hypothesis can be achieved through Zeeman line broadening measurements, although the source of chromospheric heating remains a separate issue.  Whether the larger sample of hyperactive cool dwarfs are also old, relatively massive, and possess strong magnetic fields remains to be determined, but they hint at a remarkable inversion of the standard age-activity relationship for low-mass stars.

\acknowledgments

The authors would like to thank 
Paul Sears and John Rayner at IRTF, 
Al Conrad, Heather Hersley and Marc Kassis at Keck, 
and Sergio Vera and David Osip at Las Campanas
for their assistance in the observations reported here.
Thanks also go to Shelley Wright for her help with the OSIRIS data reduction,
and our anonymous referee for her/his careful and timely critique.
This publication makes use of data 
from the Two Micron All Sky Survey, which is a
joint project of the University of Massachusetts and the Infrared
Processing and Analysis Center, and funded by the National
Aeronautics and Space Administration and the National Science Foundation.
2MASS data were obtained from the NASA/IPAC Infrared
Science Archive, which is operated by the Jet Propulsion
Laboratory, California Institute of Technology, under contract
with the National Aeronautics and Space Administration.
This research has made use of the SIMBAD database,
operated at CDS, Strasbourg, France; 
the Very-Low-Mass Binaries Archive housed at 
\url{http://www.vlmbinaries.org} and maintained by Nick Siegler, Chris Gelino, and Adam Burgasser;
and the M, L, and T dwarf compendium housed at DwarfArchives.org and 
maintained by Chris Gelino, Davy Kirkpatrick, and Adam Burgasser. 
The authors wish to recognize and acknowledge the 
very significant cultural role and reverence that 
the summit of Mauna Kea has always had within the 
indigenous Hawaiian community.  We are most fortunate 
to have the opportunity to conduct observations from this mountain.

Facilities: \facility{IRTF (SpeX); Keck (NIRC2, OSIRIS, LGS AO); Magellan Clay (MagE)}


\end{document}